\documentclass[12pt,referee]{aastex}
\usepackage[applemac]{inputenc}
\usepackage{apjfonts}
\usepackage{epsfig}
\usepackage{amsmath}
\usepackage{amssymb}
\usepackage{color}

\def\kms{\mbox{km~s$^{-1}$}}
\def\kpc{\mbox{kpc}}

\renewcommand{\AA}{\protect\hbox{$\overset{_{\circ}}{\text{A}}$}}
\newcommand{\Ha}{H$\alpha$}

\newcommand{\hi}{H{\sc I}\, }

\shorttitle{Bubble complex in NGC 6946}
\shortauthors{SANCHEZ-GIL ET AL.}

\begin{document}

\title{Kinematical and Physical properties of a 700 pc large bubble in NGC~6946}

\author{M. Carmen Sánchez Gil$^{1}$, Emilio J. Alfaro$^{1}$ and Enrique Pérez$^{1}$.}

\affil{$^{1}$ Instituto de Astrof\'{i}sica de Andaluc\'ia (CSIC), E18008, Granada, Spain}

\email{sanchezg@iaa.es}

\begin{abstract}

The galaxy NGC 6946 contains a gas-star complex of 700 pc in
diameter which appears populated by tens of young stellar
clusters, and a Super Star Cluster (SSC) as massive as 10$^6$ M$\odot$. 
The ionized gas, as drawn by the H$\alpha$ emission, delineates an almost
circular shape which we show here to be in expansion. Previous studies have
analyzed the stellar component of the complex, as well as the
structure of the atomic and ionized gas; these analyses were
restricted to the blueshifted component along the whole extent of the bubble or to a 
smaller inner region where both sides of an expanding bubble were seen. 
In this work we present a
complete spectroscopic study of this object for two position angles
crossing each other close to the young massive SSC. 
We have obtained new data with a spectral resolution six times better than previous
spectroscopic studies, taken under atmospheric conditions better
than those previously reported, allowing us to detect the approaching
and receding walls of one the largest bubbles in external galaxies
ever studied in detail.

The kinematical analysis shows a large expanding bubble, whose walls
appear to be highly structured with superposed smaller shells, likely
originated as the result of star forming events occurring at the
edges of the larger scale shell, a la Huygens. 
We also study some diagnostic diagrams of the ionized gas and conclude that
most of the observed ionization is originated by  photons from hot stars, 
but with clear evidence that some of the gas is shock ionized. 

This peculiar complex is an excellent laboratory for the analysis of the interaction 
and feedback between the gas where the stars were formed and the young and massive generation of new born stars.  

\end{abstract}

\keywords{galaxies: kinematics and dynamics, spiral, starburst --- line: profiles --- techniques: spectroscopic --- ISM:bubbles }

\section{Introduction}

The galaxy NGC\ 6946 harbors a young star cluster which can be
considered as the most massive known cluster within the limits of the
Local Volume, its estimated mass being around 10$^6$ M$\odot$,
(Larsen et al. 2001, 2006). The cluster is not isolated but it
appears to occupy the central region of a large stellar complex,
$\approx$\ 700 pc diameter, which was first reported by Hodge (1967)
and lately rediscovered and studied by Larsen \& Richtler (1999) in
their extensive mapping of young cluster populations in nearby
galaxies. Since then, this stellar complex has been extensively studied, both
through the stellar population as well as through the gas component.

 The complex contains more than a dozen of young clusters with ages
ranging between 10 to 30 million years (Elmegreen et al. 2000). A
detailed study of the stellar population performed by Larsen et
al. (2002) allowed them to derive the Star Formation History (SFH) of
the isolated stellar component, showing a present day high Star
Formation Rate (SFR) and an important decrease of the SFR of
isolated stars  at the time most of the cluster population was born. It, thus,
looks like that most of the star formation, between 10 and 15 Myr ago, took
place in the mode of cluster formation, rather than forming isolated
stars or  loose associations.

The massive cluster has an estimated age of 10-15 Myr,
 from its integrated photometry, with a luminosity of
M$_V=-13.2$ and a mass of 10$^6$ M$\odot$ as derived
from both, luminosity and dynamical criteria (Larsen et al. 2001,
2006). The internal velocity dispersion of its members has been
calculated  to be close to 9 \kms \,(Larsen et al. 2006). All these
characteristics suggest this object to be the young evolutionary stage of a
classical globular cluster. These massive young objects have been
denominated in the recent literature by different names, we adopt here
that of SSC.

The ionized gas in the complex has also been the subject of recent
studies. Efremov et al. (2002) analyzed long slit spectra, centered on
the \Ha\ emission line, and taken with the 6-m BTA and 10-m Keck-I
telescopes. The radial velocity curves along the slits show the
existence of several dips or bumps superimposed on a smooth curve
compatible with the expected velocity rotation of the galaxy at these
positions. Particularly conspicuous is the detection of a steep
fall-off in the radial velocity curve along P.A.=83$^\circ$, centered
7$\arcsec$ away from the SSC location in the side of lower emission
density, with an amplitude larger than 70 \kms. These authors
considered that feature as the signature of an expanding
semi-bubble. Three-dimensional optical spectroscopy was lately
performed with the INTEGRAL spectrograph attached to the 4-m William
Herschel telescope at Roque de los Muchachos Observatory and
with the 6-m BTA telescope at the Special Astrophysics Observatory of
the Russian Academy of Science (Alfaro et al. 2007; Alfaro \& El\'\i as 2006;
Efremov et al. 2007). The main picture emerging from these analyses is
that we are facing an \Ha\ torus, around 800 pc in diameter, which draws
an almost complete circle, 280$^\circ$ long, while it is open in the NE side of
the torus. In addition, two other \Ha\ features, one associated to
the SSC and the other forming an arc shaped structure, 130$^\circ$ long,
located at the NW of the SSC are also evident. This last feature of
\Ha\ emission is also spatially associated to the main concentration of 
the youngest stars (4 My old) in the complex ( see Fig.1 in Larsen et al. 2002) 

The atomic gas in the complex still lacks a detailed analysis at
the adequate spatial resolution, but a few important results have been
derived from some recent works (Boomsma 2007; Boomsma et al. 2008;
Efremov et al. 2007). The complex appears to be placed in a region of
 \hi low density, which extends further to the NE, forming an
extensive \hi hole a few \kpc\ in size. This is not strange since NGC\
6946 is known to be a galaxy with a large number of \hi holes and High
Velocity Clouds  (HVC) (Kamphuis 1993), some of which have kinetic energies
above 10$^{53}$ erg.  A look to Fig. 14 in Efremov et
al. (2007) shows that, at the position of the SSC, the distribution of
\hi density presents a local minimum for a wide range of gas
velocities, and some smaller knots are visible at velocities well below
the mean value.

All these features make  this region an excellent lab for analyzing
the interaction and feedback between the gas and the youngest
products of the star formation processes, in particular, concerning
the small-scale relationship between the interstellar medium (ISM)
and the SSC.  Good spectral and spatial resolution are thus crucial
for determining the main physical properties of the gas along the
entire super-bubble and for drawing the true geometry of the
complex. Through the data presented here, we attempt to clarify which
physical mechanisms could apply in this region of NGC\ 6946. 

The paper is organized into 6 sections; the first is this
introduction. In section 2 we describe the observations and the
primary reduction of the spectra. The analysis of the emission lines
is shown in section 3, and in sections 4 and 5 we explain and discuss the  
results obtained. Finally, section 6 summarizes the main conclusions of this work.

\section{Observations and data reduction}\label{Sec1}

We obtained long-slit spectroscopy with the double arm ISIS spectrograph attached to the  4.2m William Herschel Telescope (WHT), at the Roque de los Muchachos Observatory (La Palma) during August 2003. This instrument consists of two intermediate dispersion spectrographs operating simultaneously, separating the blue and red spectral regions, which are imaged onto an EEV12 and a  MARCONI2 CCD chips respectively. In this way we have two spectral ranges observed simultaneously, a blue one centered around H$\beta$ (4861 \AA), and a red one around \Ha\ (6563 \AA).  The gratings used, R1200R  and R1200B, provide a dispersion of 0.23 \AA/pixel. The slit width of 1 arcsec projects onto about 3.64 pixels Full-Width-Half-Maximum (FWHM) on the detector; the spatial sampling along the slit is 0.2 arcsec/pixel. The slit was placed at two position angles, 0$^\circ$ and 296$^\circ$, approximately centred on the shell (Fig. 1). In  table \ref{table1} we present a comparison of our observational configuration
with previous spectroscopic studies dealing with this object and in table \ref{table2} the log of the observations.

The spectra were reduced and calibrated following the standard procedure. Bias subtraction, flat-fielding, wavelength and flux calibration were done with the IRAF\footnote{IRAF is distributed by the National Optical Astronomy Observatory, which is operated by the Association of Universities for Research in Astronomy (AURA) under cooperative agreement with the National Science Foundation} task {\tt ccdproc}. For the wavelength calibration CuNe and CuAr lamps were used. The standard star Feige 110 from the Oke (1990) catalogue was used for flux calibration. Sky subtraction was done using the IRAF NOAO package task {\tt background}. For each wavelength range, the different exposures at each slit orientation were combined, eliminating cosmic rays and bad pixels, thus obtaining four final spectra. 

We use an intermediate band \Ha\ image (central wavelength 657.7 nm and FWHM=18 nm) taken with the Nordic Optical telescope (NOT)\footnote{The image kindly made available by Y. Efremov is taken through a filter called wide '\Ha\ filter' in the NOT system}.

\section{Spectral Analysis: Emission Line Fitting}\label{Sec2}

The final combined sky subtracted red spectrum at P.A.=0$^{\circ}$ is shown in Fig. 2a, in the spectral range 6540-6750 \AA\ covering the five main emission lines, \Ha, [NII]6548,6583, and [SII]6716,6731. Fig 2b shows an expanded view around \Ha\ of the spectrum along P.A.=0$^{\circ}$, 296$^{\circ}$ (not  sky subtracted). The origin of the spatial scale has been arbitrarily set to a point close to the shell center (where the two slits intersect). The residuals of the sky subtraction can be seen (in the top panel) along the slit at 6553.5 and 6562.8 \AA. Upon close visual inspection these spectra readily show that the \Ha\ emission is highly structured in velocity along the slit, with a series of components that indicate the presence of shells. In this section we describe the detailed measurement of these structures with the aim of obtaining the kinematical information along the two slit positions.

We fit the \Ha\ emission line at each pixel along the slit. Line flux, FWHM, {\Large }central wavelength, and their corresponding errors, are calculated with the STARLINK\footnote{http://star-www.rl.ac.uk/star/docs/sun50.htx/sun50.html} package DIPSO. The continuum level is simultaneously fitted with a first order polynomial. 

The  sky subtracted spectra in Fig.2 (top panel) show residuals in some sky emission lines.  Particularly, the atmospheric \Ha\ line, is relatively conspicuous and close to the \Ha\ emission in NGC 6946. 
After some trials we decided to fit the spectra prior to the sky subtraction (bottom panel). The reason is that, at some locations along the slit, there is a faint high velocity blue shifted \Ha\ component very close to the sky \Ha\ emission, and the fitting subroutines work better by fitting an additional component to the sky emission, while the results are worst working with the residual sky emission in the sky subtracted spectra. This method has the additional advantage of obtaining a map of the uniformity of the spectral resolution and quality of the wavelength calibration along the slit, from the fits to the sky emission lines. 

Figure \ref{fig:f2} illustrates two example fits of the slit at P.A.=0$^\circ$. In the case of the spatial increment 500 along the slit, we fit two components to the \Ha\ nebular emission, and one component to each of the sky emission lines at 6562.8 \AA \, and 6568.9 \AA, respectively. The case of the spatial increment 565 along the slit is an example where we could fit three components to the nebular \Ha. This extra third component is present only in some locations along the slit, while in most other locations the fit is satisfactory with just two components (inside the stellar complex) or with just a single one (corresponding to the emission outside the complex, see next section for details). 
From these fits we estimate the values for the following variables,

\begin{eqnarray}
V_{obs} \quad &=& \quad \frac{(\lambda_{obs}-\lambda_0)\, c}{  \lambda_0} \label{eq:1}\\
\nonumber \\
FWHM \quad &=&  \quad W_{obs} \frac{c}{ \lambda_0 } \label{eq:2}\\
\nonumber \\
EW  \quad &=&  \quad \frac{line\, intensity }{I_c } \label{eq:3}\\
\nonumber \\
r  \quad &=&  \quad (y - y_0)\, 0.2\, \frac{1}{3600} \, \frac{\pi}{180} \, 5.5\times10^6 \label{eq:4}
\end{eqnarray}

\noindent where {\it V$_{obs}$} is the line velocity in \kms, observed from the Earth, {\it c} is the velocity of light and $\lambda_0 \, =$ 6562.8 \AA\ the air reference wavelength of the line. Observed velocities were then tied to the heliocentric system. So the velocity of the gas, throughout this paper, refers to the heliocentric velocity. In equation (\ref{eq:2}) $FWHM$ is expressed in \kms, $W_{obs}$ is the FWHM from the fit. In equation (\ref{eq:3}) $EW$ is the equivalent width and $I_c$ the continuum flux, which corresponds with the independent term of the first order polynomial fit. The {\it line intensity} is the integrated flux of the corresponding gaussian component. Finally, in equation (\ref{eq:4}), {\it r} is the relative distance, in pc, from a fixed point of the spectrum {\it y$_0$}. Such point corresponds with the intersection of the two slits (in each case); {\it y} is the pixel coordinate along the slit, where the scale is 0.2 arcseconds per pixel and 5.5 Mpc is the distance to NGC 6946 (Sofue 1997).

\section{Results}\label{Sec3}

Previous studies of this star forming complex suggest that we are facing a large superbubble in expansion (Efremov 1999; Elmegreen et al. 2000; Efremov et al. 2002, 2007; Larsen et al. 2002). Here we perform a detailed analysis on the \Ha\ kinematics and how it relates with the physical properties and geometry of the emission gas. 

The most significant result from the \Ha\ fitting is that the emission line is resolved into (at least) two kinematic components of the gas along the full extent of the complex, $\sim700$ pc. We refer as  the 'blue component' of the gas that  moving towards the observer at velocities of up to $\sim50$ km s$^{-1}$, and as  the 'red component',  the one receding at similar velocities.

Table \ref{table3} contains a summary of the mean values for the two kinematic components along the extent of the bubble, this is $\sim$(-21, 29) arcsec, or $\sim$(-660, 773) pc equivalently, from the bubble center for P.A. = 0$^\circ$, and $\sim$(-52, 28) arcsec, $\sim$(-1387, 747) pc, for P.A. = 296$^\circ$. At P.A. = 0$^\circ$, the mean velocity is 106 \kms\ for the blue and 148 \kms\ for the red component, while at  P.A. = 296$^\circ$ it is 120 \kms\ and 169 \kms, respectively.
Outside this area we could identify only one velocity component for \Ha, which corresponds to  the rotation velocity of the gas disk at the super complex location. The different line components observed is a clear signature of a bubble kinematics, and the radial velocities measured provide a lower limit to the expansion speed. We study this in detail in sub-section \ref{Sec3b1}, once we have determined the main geometric parameters of the complex.

\subsection{Kinematics and Geometry: Expanding bubbles}\label{Sec3a}

An analysis of the kinematics and geometry of the ionized gas is also derived  from the results of the fits, which are summarized graphically in figure \ref{fig:f3}. The heliocentric velocities  are plotted  versus the relative distance for all the positions along the slit. The reference position for the spatial scale corresponds to the intersection of the two slits (cf. fig.\ref{fig:f8}). The red component of the gas is represented with filled triangles (red color in the electronic version), and the blue component with filled circles ( blue color).
We can observe in these plots a complex structure of the gas, having several components with different kinematics. The overall kinematic structure is that of a gas bubble in expansion, sufficiently transparent so that the approaching blue component does not obscure the receding red component. The high spatial and spectral sampling of these data allow us  to disentangle the expanding kinematics of the bubble.

For the case of P.A. = 0$^\circ$, we could fit up to three kinematic components to \Ha\  in some regions of the spectrum. This third component appears blueshifted at high velocities ($\sim$85 \kms) with respect to to the rotation velocity of the galaxy at its respective position. It is represented as filled squares (and light blue) in figure \ref{fig:f3}.

For the spectrum at P.A. = 296$^\circ$ we always found two components as the best fit to the nebular \Ha\ emission along the whole extent of the complex.

Efremov et al. (2002) studied the kinematics of the ionized gas within this complex through the \Ha\ emission, 
using long-slit spectroscopy at several positions with the 6m BTA telescope. Only  the blueward side of the bubble was detected and analyzed by these authors. The lower spectral dispersion (six times lower) and the larger seeing of their data as compared with the data we present here may be the reasons why Efremov et al. (2002) did detect only one main component.
 A 14\arcsec\ long slit centered at the SSC, at P.A. =$-10^\circ$, taken with the Keck-I telescope (see table \ref{table1}), shows both velocity components in emission extending on both sides of the SSC. It is in this spectrum where they detect a bubble 6\arcsec\  in radius and 60 \kms\ expansion.  

The rest of the spectrum along the two slit position in our data shows the \Ha\ emission of the interstellar medium surrounding the complex, and only one component was fitted. Emission from other \Ha\ regions to the north and west ends of the slits can be seen in figures \ref{fig:f4} and \ref{fig:f6}. In figure \ref{fig:f6} the slits are represented on a \Ha\ image of the galaxy. Figure \ref{fig:f4} shows the complete kinematic map of \Ha\ along P.A. = 0$^\circ$, together with the rotation curve of the galaxy described in section \ref{Sec3b} below (bold black line with its error limits in gray).  The north tip of the spectrum could be  fitted with two components but this part of the data set are out of the scope of this work, so we do not further analyze them. 

The bubble is clearly seen in figure \ref{fig:f4}  between -500 pc and 400 pc with the kinematic components well differentiated. Outside the bubble the emission appears to fit the gas rotation velocity at this galactic location (blue points); also an approaching component, at $\sim60$ \kms, is present along the whole extent of the slit. The origin of the bluest component is still controversial, and it could represent the diffuse ionized gas (DIG) in the halo of the host galaxy (NGC~6946). Its constant velocity and
low emissivity along the whole extent of the slit support this hypothesis, 
because the extraplanar gas does not follow the rotation curve of the disk
(Heald et al. 2006). This velocity feature is also found by Efremov et al (2007) in the HI emission. At the top right-hand and bottom left-hand panels of their figure 14, the HI position-velocity diagrams for two slits are represented, with different position angles, and the structure in velocity is very similar to what we find here.  A more detailed study is needed to elucidate the origin of this emission. 

Errors in velocity measurements are larger for fainter intensities. In order to obtain reliable estimations of the bubble expansion velocities and to contrast the complex structure of the ionized gas, a filter is applied to eliminate those points with worse signal-to-noise ratio. We take as detection limit the weakest sky emission line that we can measure in the 2D spectrum. Since the sky line flux must be constant along the slit, its variation gives us an approximation of the dispersion in the measurement of the faintest emission lines. Moreover, the intensity of the \Ha\ emission in the disk of NGC 6946 outside the complex coincides with that of the weaker sky line, so the fluctuations of these lines about their mean value provide a good indicator of the uncertainty in measuring the weakest lines. The reference dispersion of the flux, $\sigma_f$, that we measure is $\sigma_f=1.2\times10^{-17}$ erg cm$^{-2}$ s$^{-1}$.

Most of the points in the top panels of figure \ref{fig:f3} have measured fluxes $>3\sigma_f$. The bottom panels represent just the same data but only for those points where the flux is  $>9\sigma_f$. This is a very stringent criterium, but the result tells about the reliability of the kinematic structures seen in the data. If we compare with the top of figure \ref{fig:f3}, now the structure is somewhat cleaner, but all the important features remain of the expanding bubble with much substructure around. Only some of the very faint points of the extremely blueshifted component remain after this strong constraint, although all of them have fluxes $>3\sigma_f$.

Furthermore, at bottom panel of figure \ref{fig:f3} there is a gap around a relative distance of -200 pc, more noticeable at P.A. = 296$^\circ$. These gaps in the kinematics are clearly visible in the 2D-spectrum in figure \ref{fig:f1}, not only for \Ha\ but also for [NII] and [SII]. The HI data in Efremov et al. (2007) also show this gap in the position-velocity diagram (cf. their figures 14, 19 and 21). 

Dashed lines, in the bottom panels of figure \ref{fig:f3} mark approximately the maximum expansion velocities of both components. So, along P.A. = 296$^\circ$ we find a radial velocity amplitude of 60 \kms, corresponding to the larger difference in the main structure.  Overimposed to the large bubble we can observe smaller structures at each side of the large shell, which show deviations of $\sim35$ \kms\ with respect to the main velocity of the bubble walls. At P.A. = 0$^\circ$ the main radial velocity amplitude is 45 \kms, while the velocity deviations of the smaller substructures are always below 25  \kms. 

The solid lines that run through the complex represent the rotation curve of the galaxy NGC 6946, corresponding to the complex zone. More details about how it has been calculated is described in next subsection. The rotation curve sets the systemic velocity of the galaxy around the complex, and it does indeed matches the mean velocity computed for the expanding gas. These velocities are around 135 and 140 \kms, somewhat larger than the value given by Bonnarel et al. (1986, 1988) of around 125 \kms.

To better visualize our model of the gas kinematic structure, the top panel in figure \ref{fig:f5} shows a simple sketch drawn on the velocity-distance diagram. This figure is similar to fig.  \ref{fig:f3}, but now the velocity is in the ordinate axis and the points are drawn with their error bar. The figure also includes the corresponding color coded fluxes of the different components (units on the right hand side ordinate axis). This panel also includes a sub-panel with the instrument-corrected velocity dispersion of the main blue component along the slit.
The bottom detached panel plots the velocity dispersion of the two main components versus the \Ha\ flux for all the points in the complex along the slit; its analysis is described below, at the end of the section. First, we compare and analyse the model at the top with the results from figure \ref{fig:f3}.

This sketch gives an idea of the large size of the gas structure in expansion, the main bubble. As apparent from a cursory look, the main kinematic shell is highly structured. Both sides of the shell (approaching and receding), along the two slit positions, contain localized structures that could be interpreted as superposed smaller shells, likely having been formed as the result of star forming events occurring at the edges of the larger scale shell (Larsen et al. 2002). 

These smaller shells or bubbles detected at some locations along the two sides of the main shell of the complex do not appear as complete shells, but rather we only see one of their sides.
One example of a small bubble is drawn in the receding main component of the bubble complex at location -200 pc, where we would be measuring a small bubble but only its receding component at $\sim$30 \kms\ with respect to the main receding kinematic component at 170 \kms. Two additional examples are drawn on approaching component of  the main complex; we have sketched two small bubbles at 0 pc and -250 pc where we can measure their receding component at $\sim$20 \kms\ with respect to the main approaching kinematic component at 110 \kms.

\begin{figure*}
\begin{center}
\includegraphics[width=\textwidth,height=\textwidth]{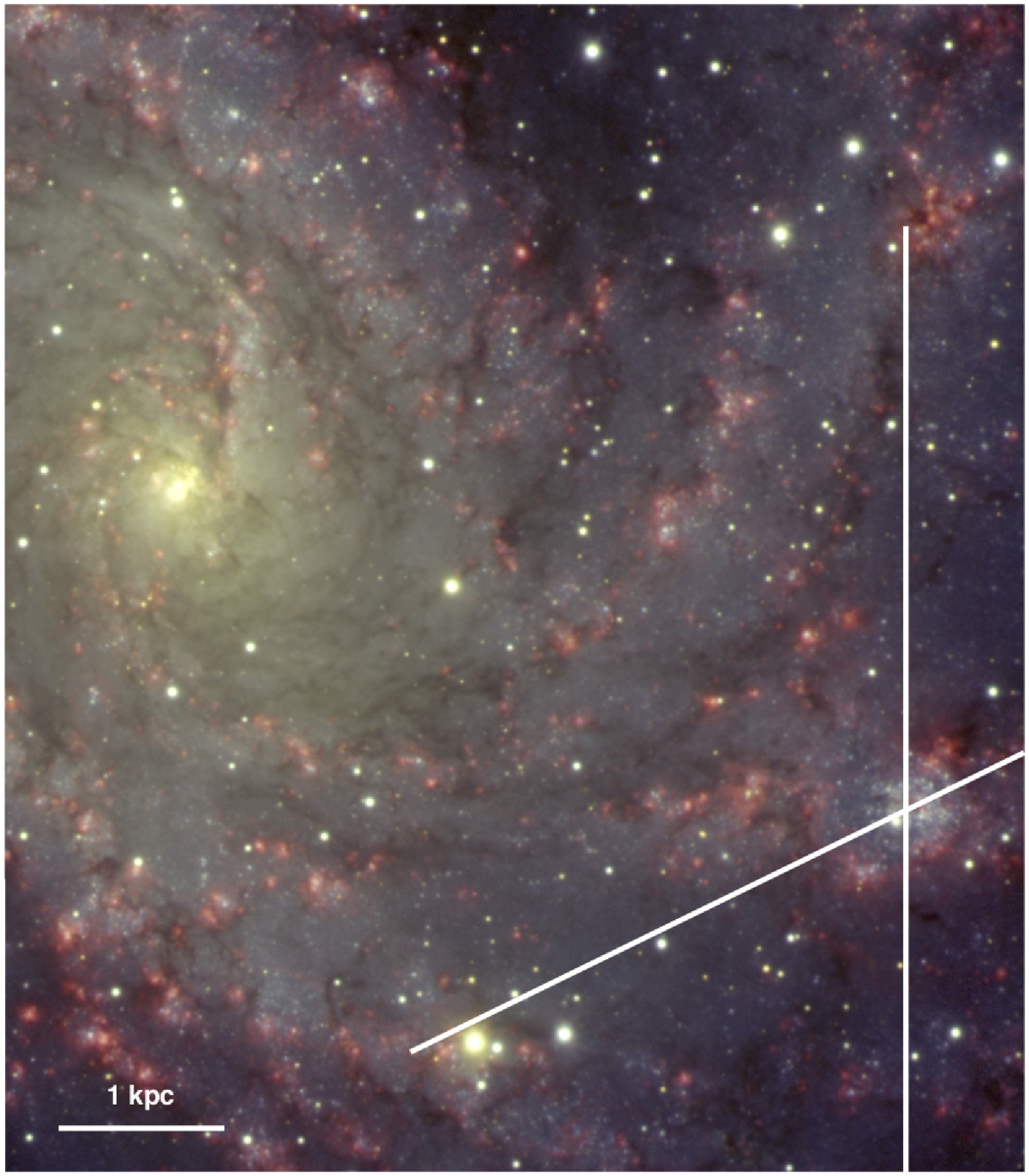}
\caption{\footnotesize Gemini Multi-Object Spectrograph (GMOS) image from Gemini North on Mauna Kea, in the bands g', r', i', and \Ha\ (Gemini Observatory / Travis Rector, University of Alaska Anchorage). The image has been adapted, showing the relative position of the complex with respect to the galaxy center. The two slit positions used in this study are also drawn on the complex. North is up and East to the left.}
\label{fig:f0}
\end{center}
\end{figure*}

\begin{figure*}
\begin{center}
\includegraphics[width=0.8\textwidth]{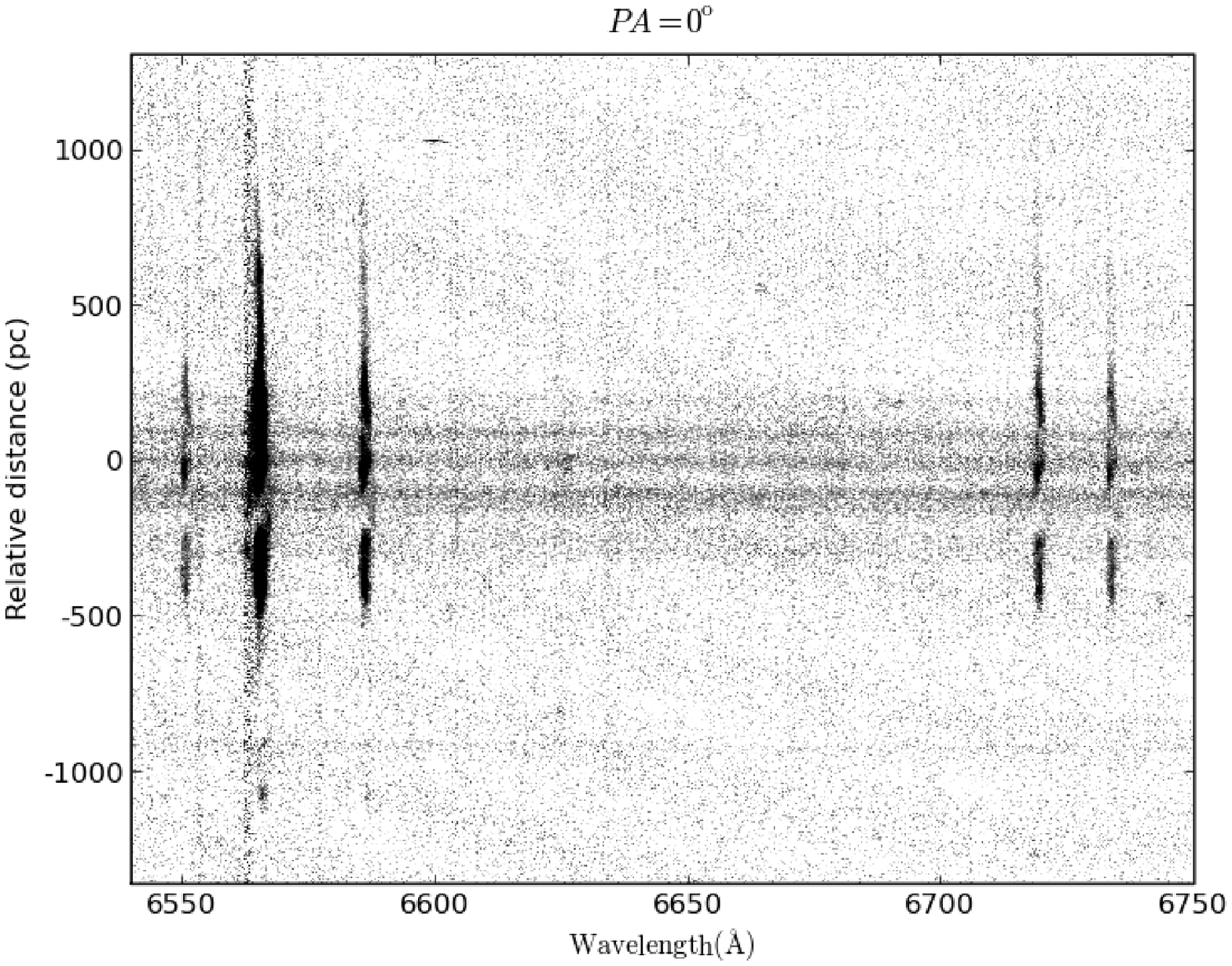}
\includegraphics[width=0.45\textwidth]{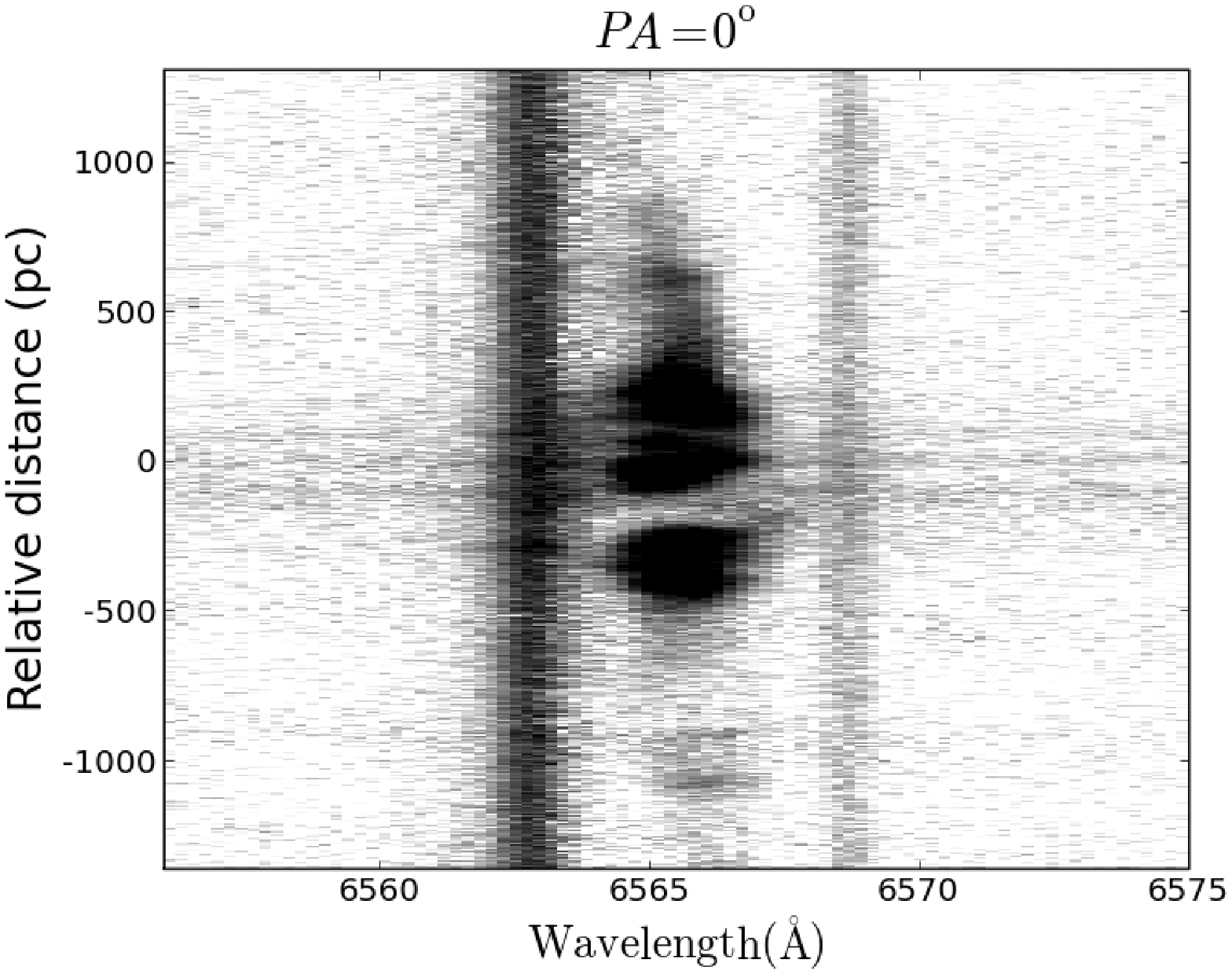}
\includegraphics[width=0.45\textwidth]{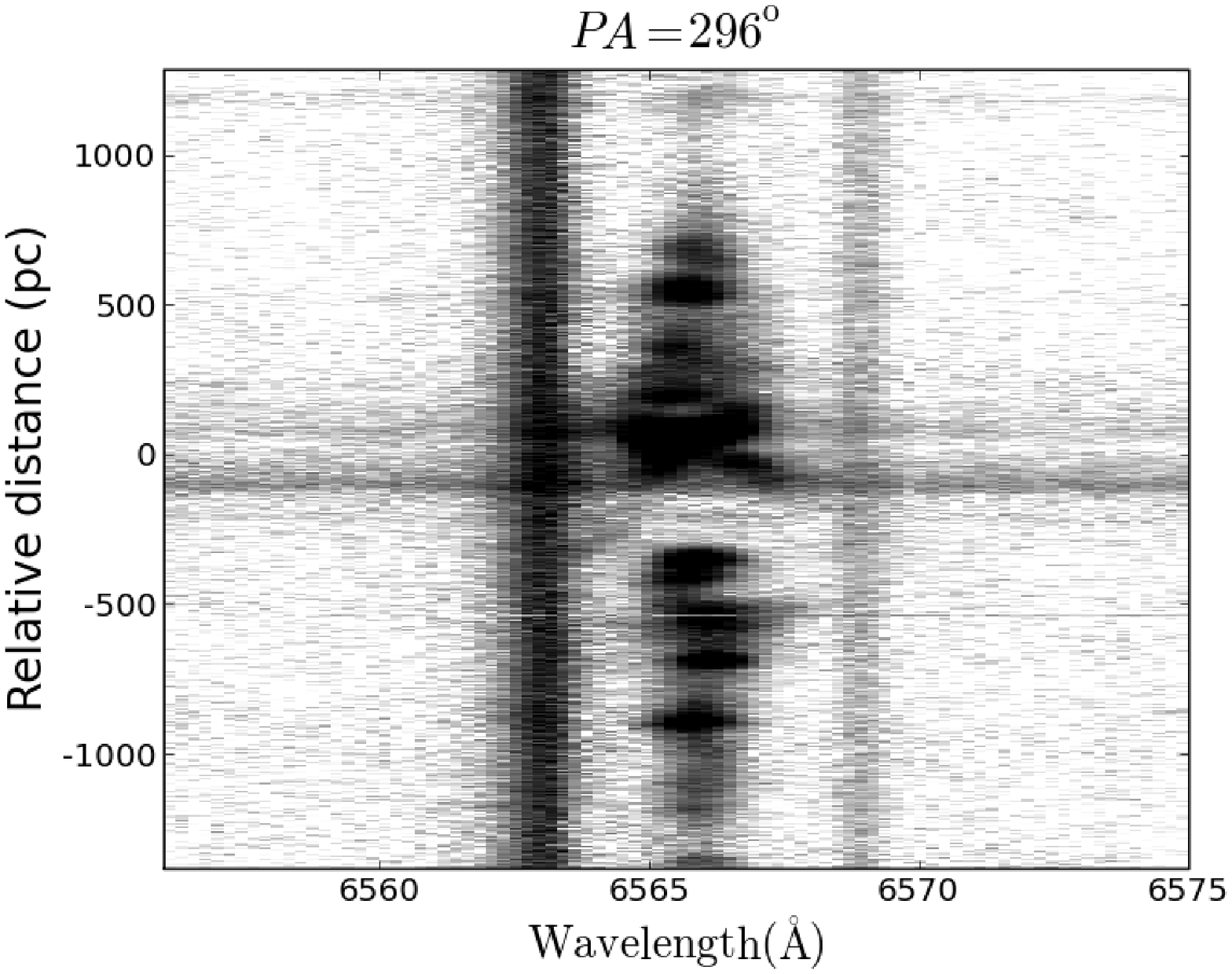}
\end{center}
\caption{\footnotesize The 2D long-slit red spectra, centered at \Ha, for both slit positions; relative distance increases towards the North. Top, the sky subtracted red spectrum for P.A. = 0$^\circ$, showing the \Ha, [NII] $\lambda$6548, 6584, and [SII]  $\lambda$6716, 6730A, emission lines. 
Bottom, red spectra (not sky subtracted) for both slit positions zoomed into \Ha, show a detailed complex structure of the ionized gas. } 
\label{fig:f1}
\end{figure*}

\begin{figure*}
\centering
\includegraphics[width=0.45\textwidth]{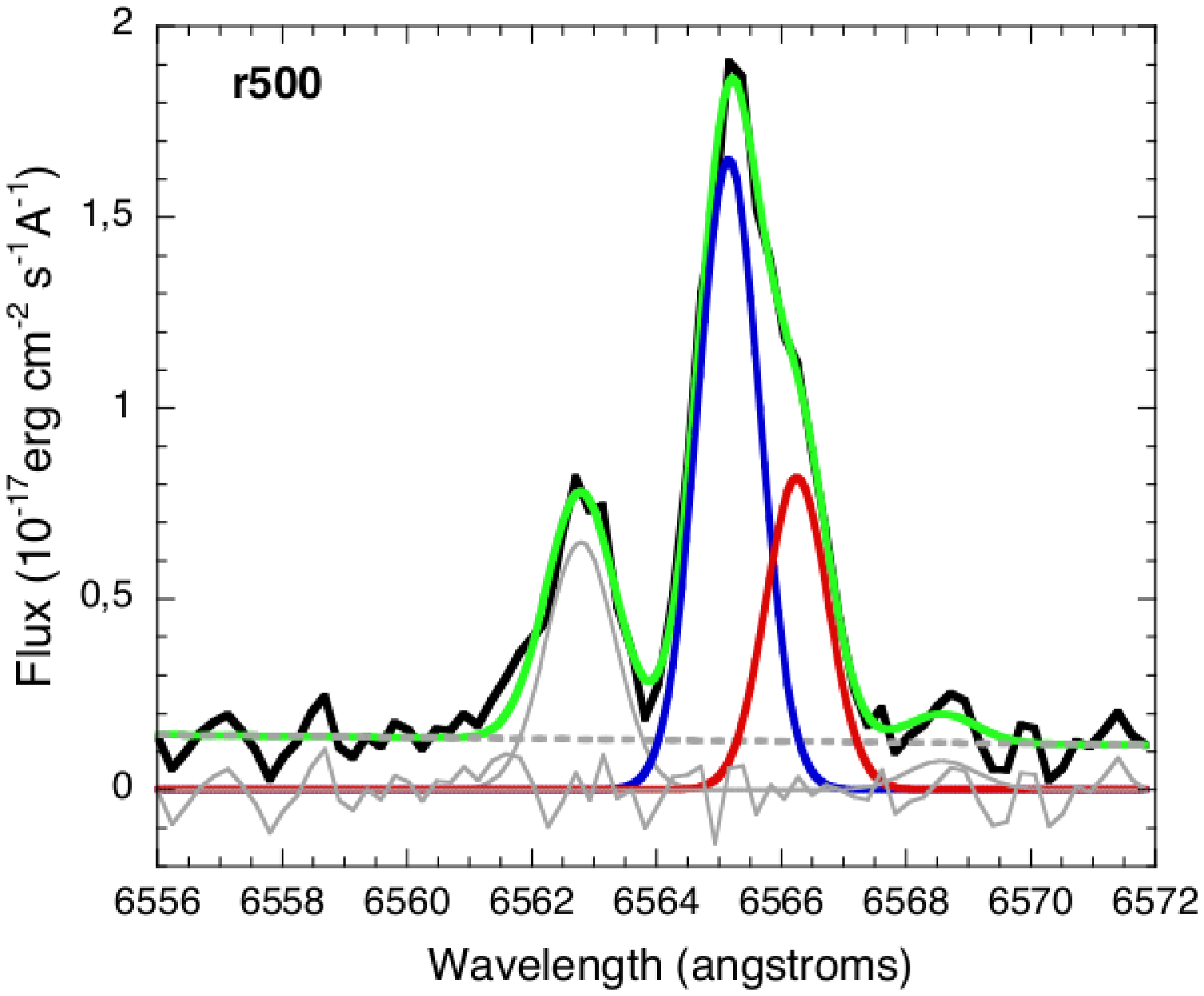} 
\includegraphics[width=0.45\textwidth]{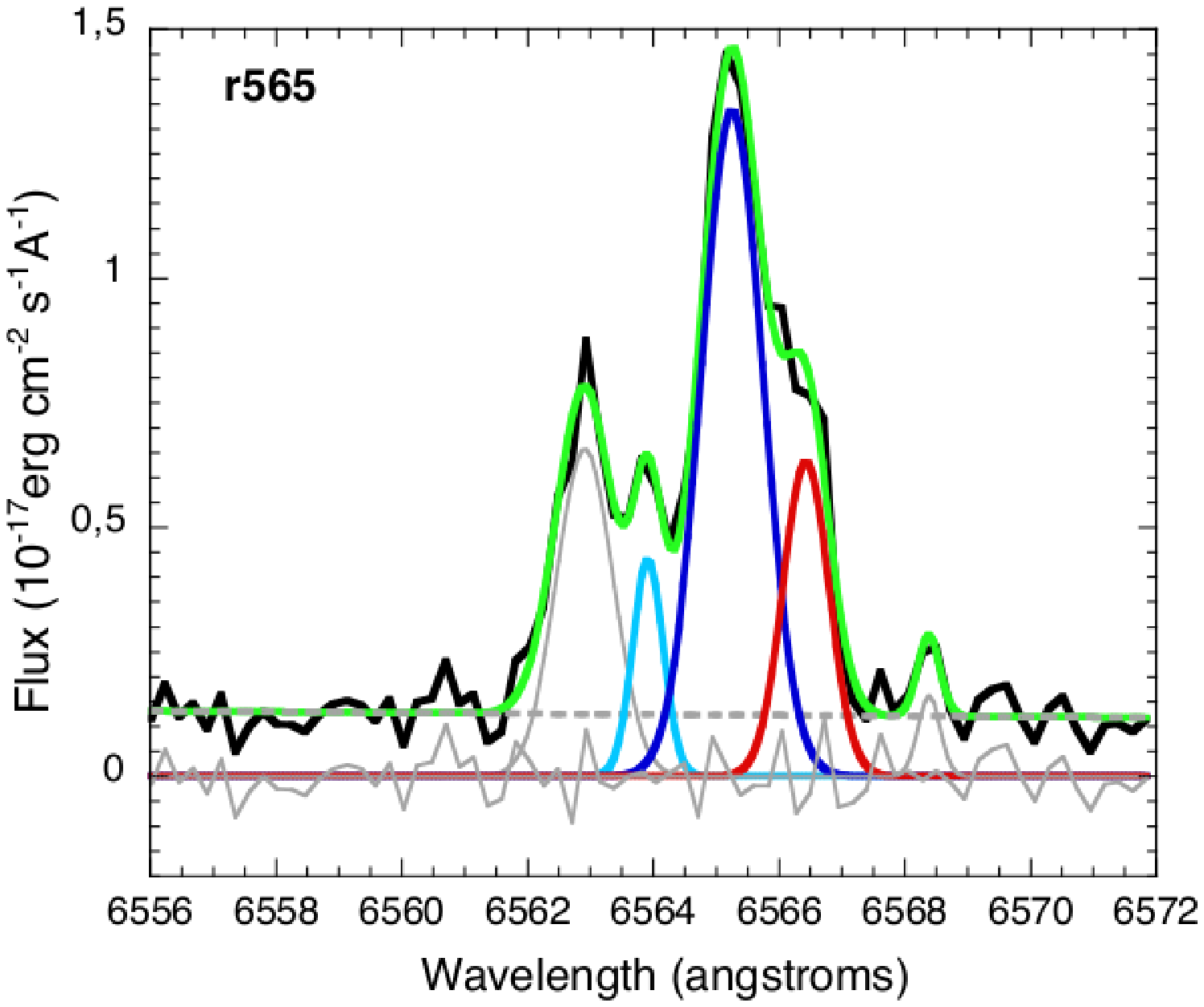}
\caption{\footnotesize Examples of two different fit cases. Left (r500), fit with two components for \Ha\ and two for the sky. Right (r565), three components for \Ha\ and one for the sky. These components are located at the center and represented with bold lines, with the same color-code that in figure \ref{fig:f3}: red for the 'red' component, receding gas, and blue for the 'blue' component, approaching gas. Light blue is for the third component, blueshifted at higher velocities. The component in grey corresponds to the sky background. The dashed line is the continuum fit. The green bold line is the global fit for the spectrum (black, bold line)}
\label{fig:f2}
\end{figure*}

\begin{figure*}
\begin{center}
\includegraphics[width=0.45\textwidth,height=8cm]{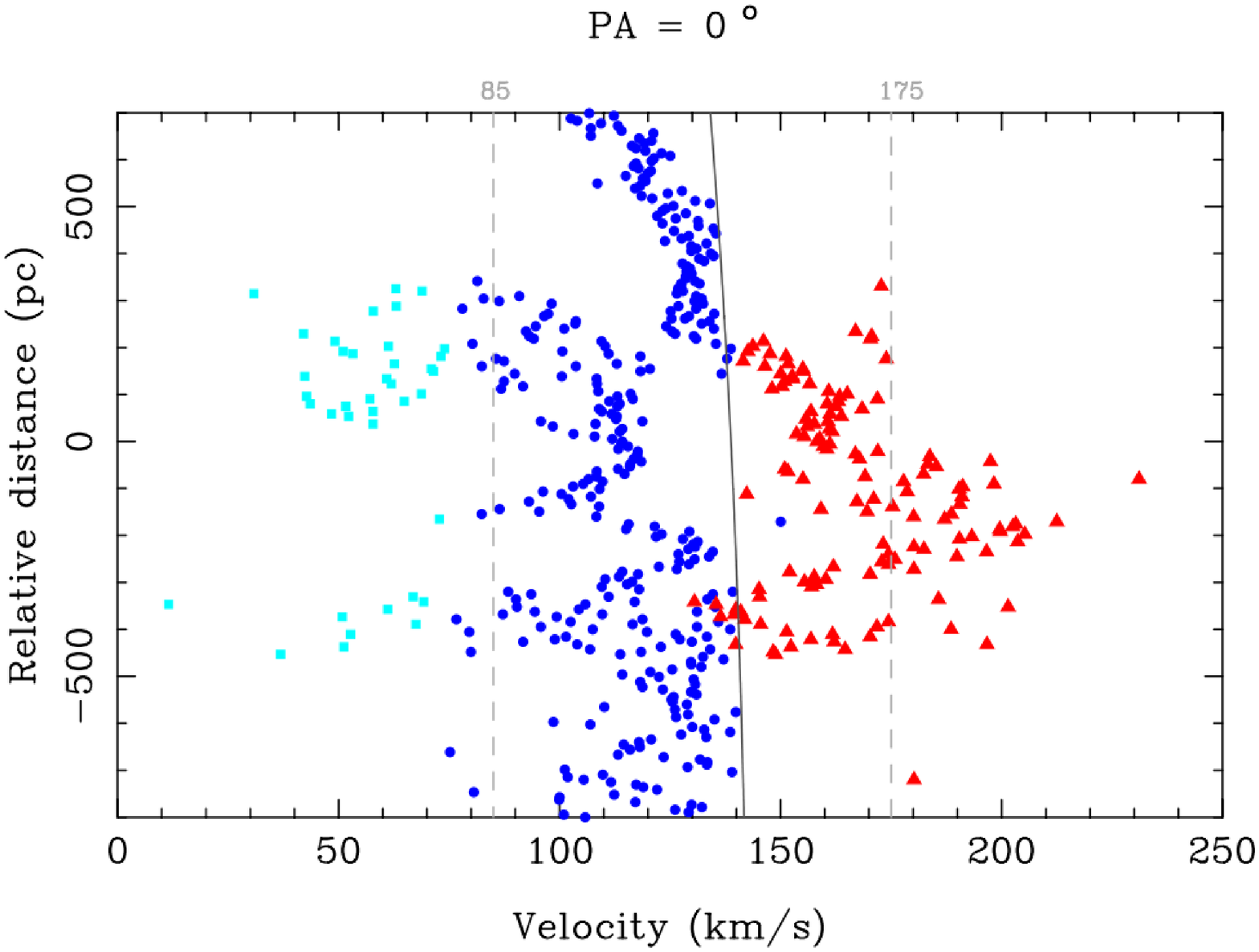}
\includegraphics[width=0.45\textwidth,height=8cm]{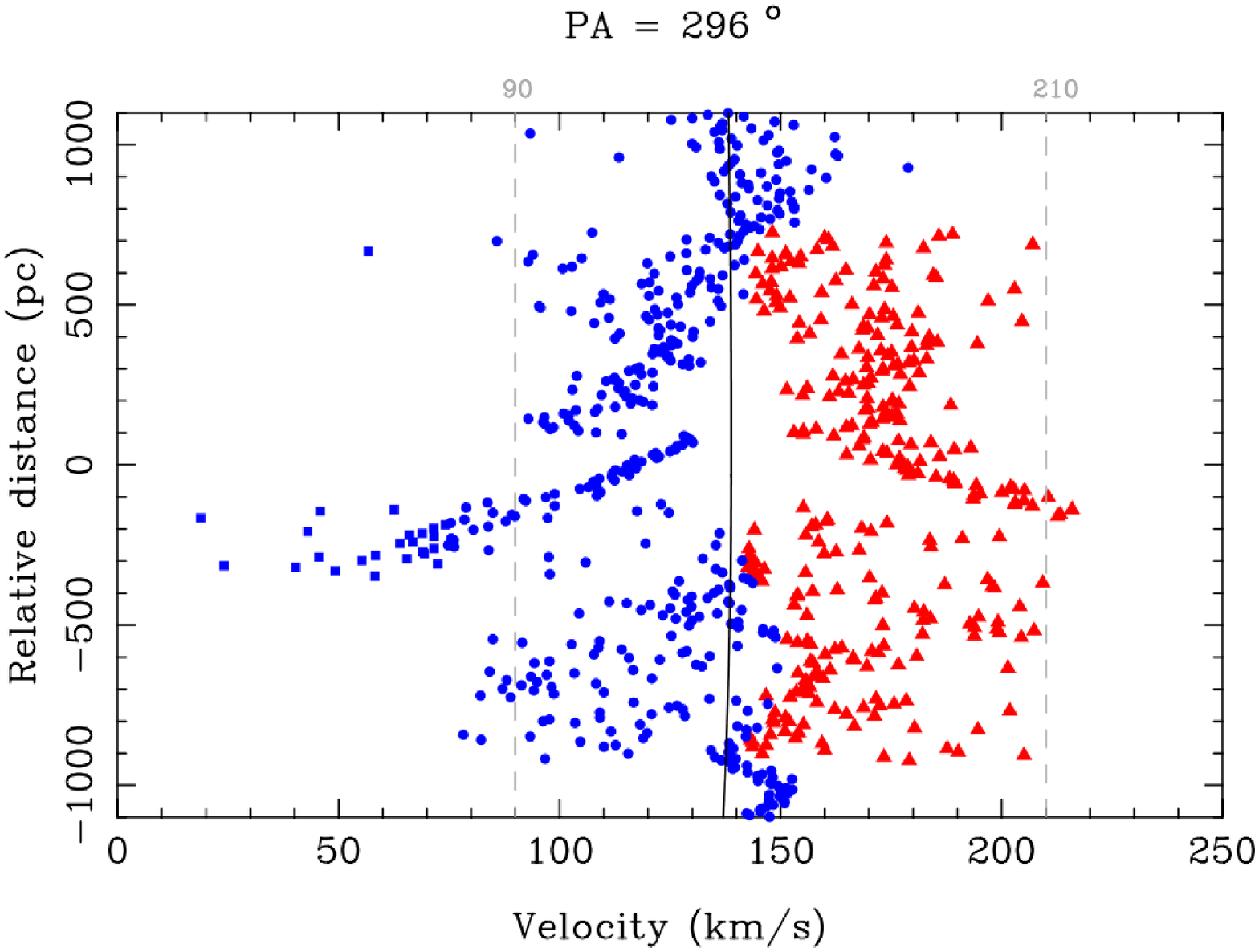}
\includegraphics[width=0.45\textwidth,height=8cm]{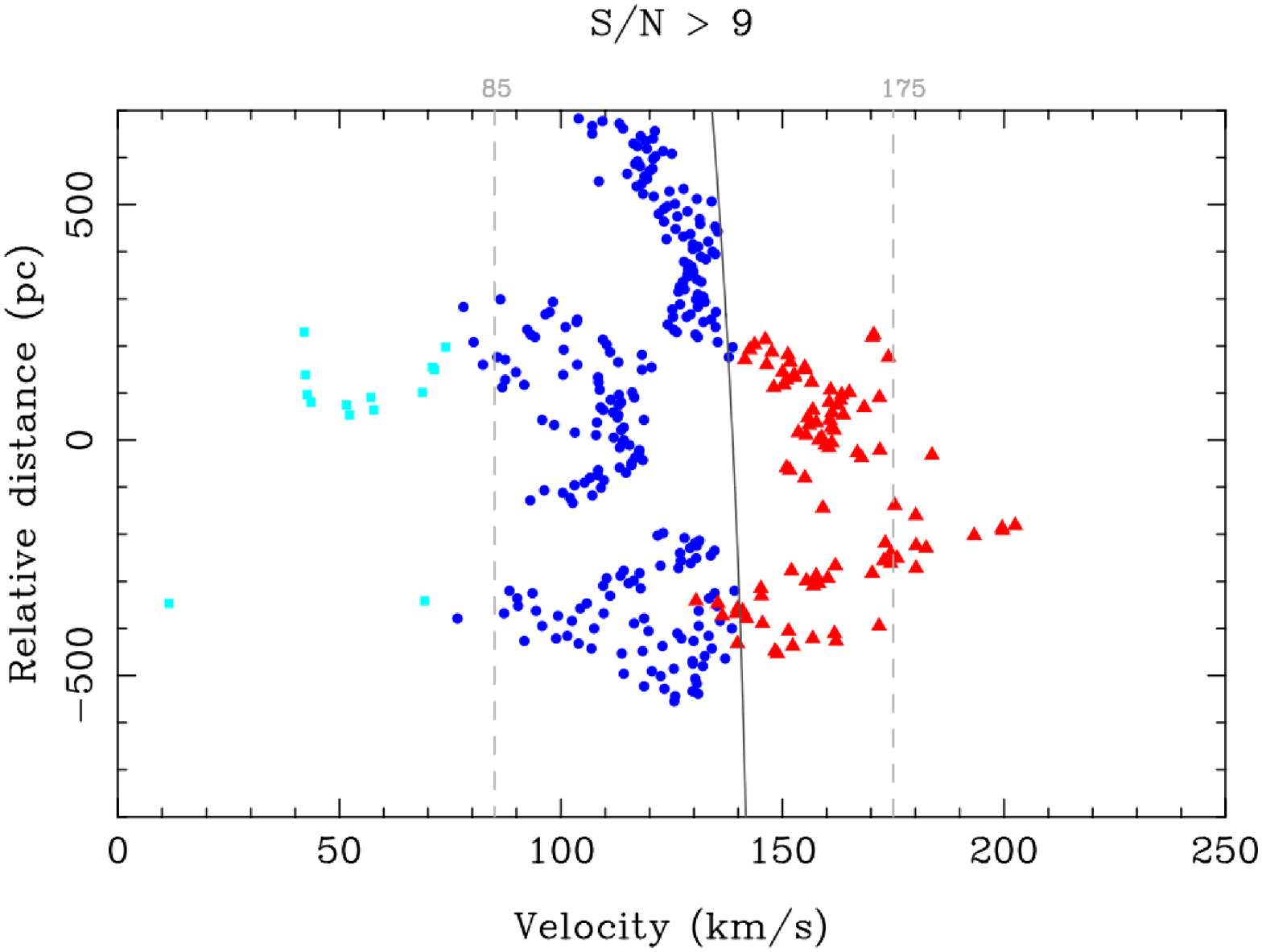}
\includegraphics[width=0.45\textwidth,height=8cm]{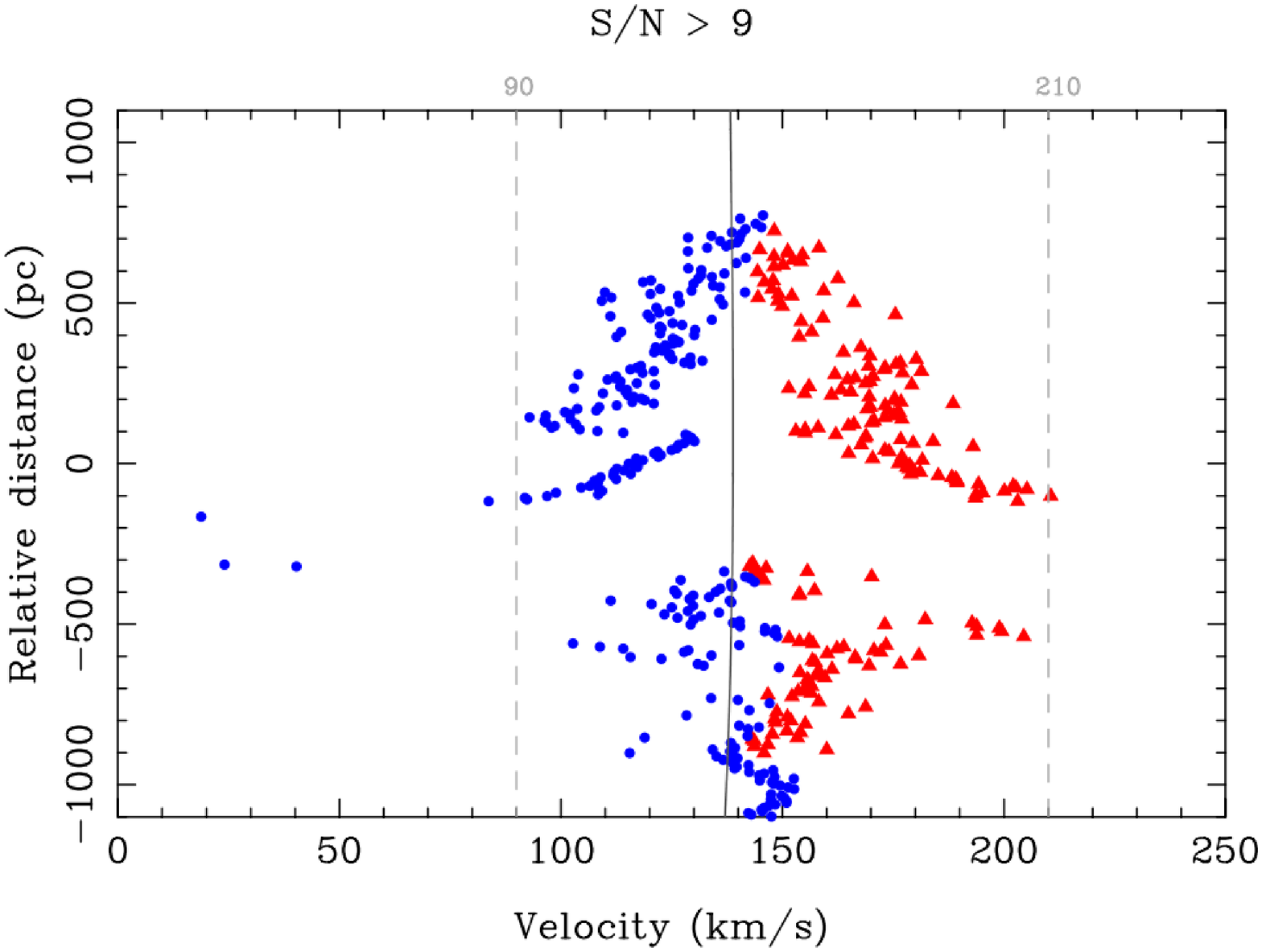}
\end{center}
\caption{\footnotesize Top left: \Ha\ velocity along the slit at P.A.=0$^\circ$.  In some locations only one component could be fitted (filled blue circles), in other locations two components were present (blue circles for the approaching and red triangles for the receding), while at other locations a third component at higher approaching velocities could be clearly identified (light blue squares). Top right: the same for the slit at P.A. = 296$^\circ$; a maximum of two components could clearly be fitted. The bottom panels show the same data but after removing those points with fluxes smaller than 9$\sigma_f$, where $\sigma_f$ is the uncertainty in measuring the faintest emission line in the 2D spectrum (see text for details). }
\label{fig:f3}
\end{figure*}

\begin{figure}
 \begin{center}
\includegraphics[width=0.5\textwidth]{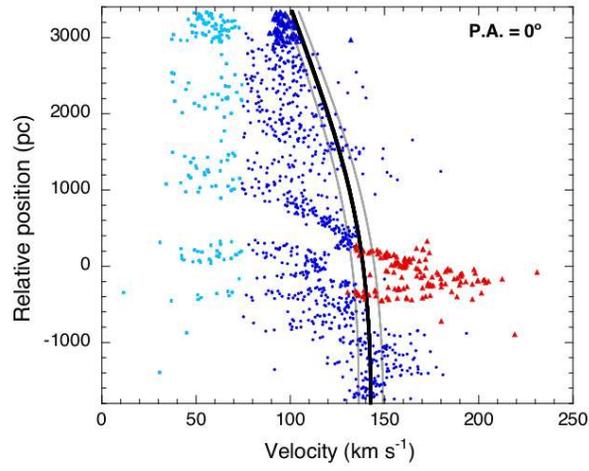}
\end{center}
\caption{\footnotesize Same plot as in figure \ref{fig:f3}, for the case P.A. = 0$^\circ$, along 5330 pc representing nearly the full extent of the slit, where it was possible to fit at least one component to \Ha. The rotation curve of NGC 6946 is also represented with the error limits from the fit (more details in section \ref{Sec3b}). }
\label{fig:f4}
\end{figure}

\begin{figure}
\begin{center}
\includegraphics[width=0.5\textwidth]{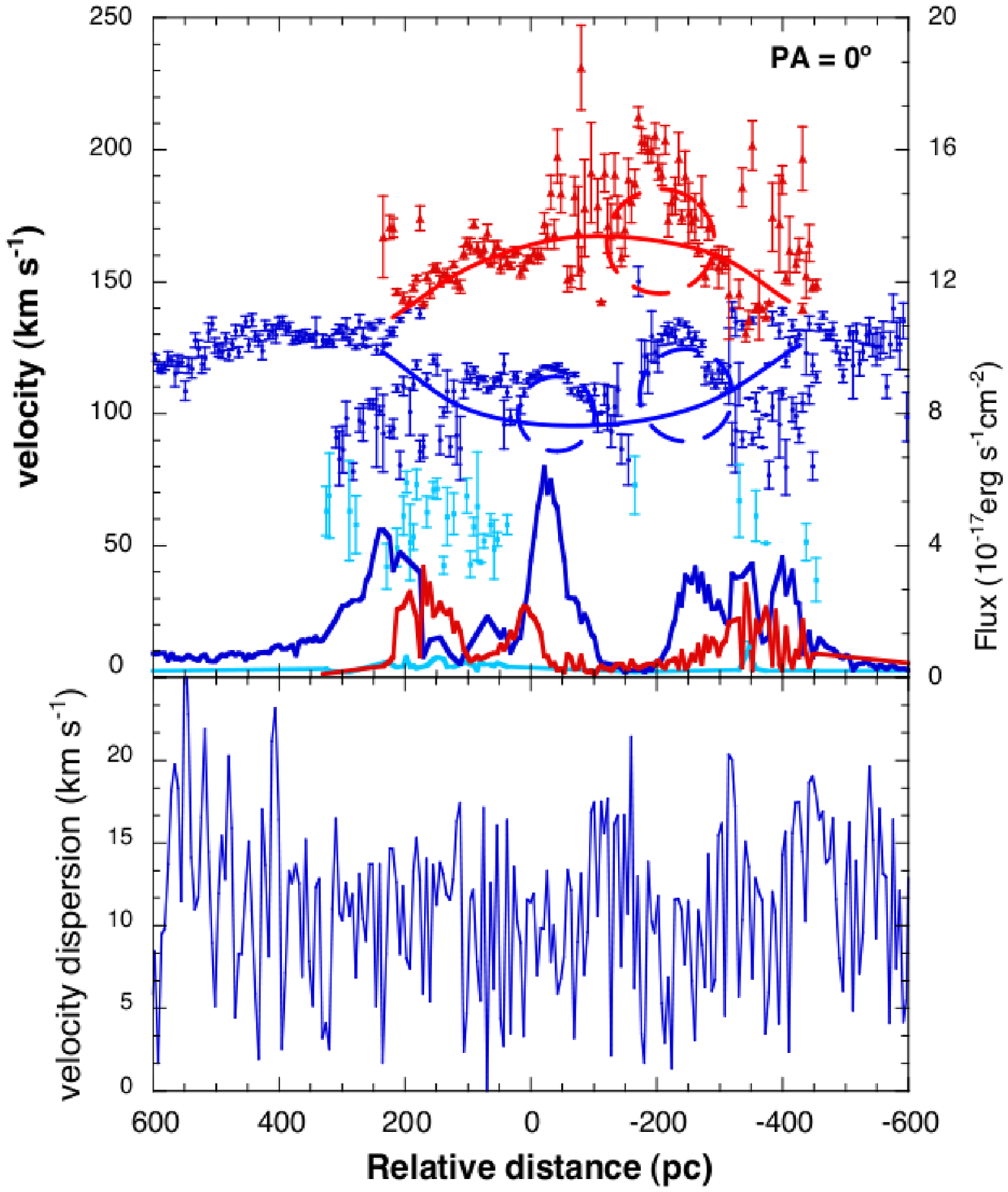}
\includegraphics[width=0.5\textwidth]{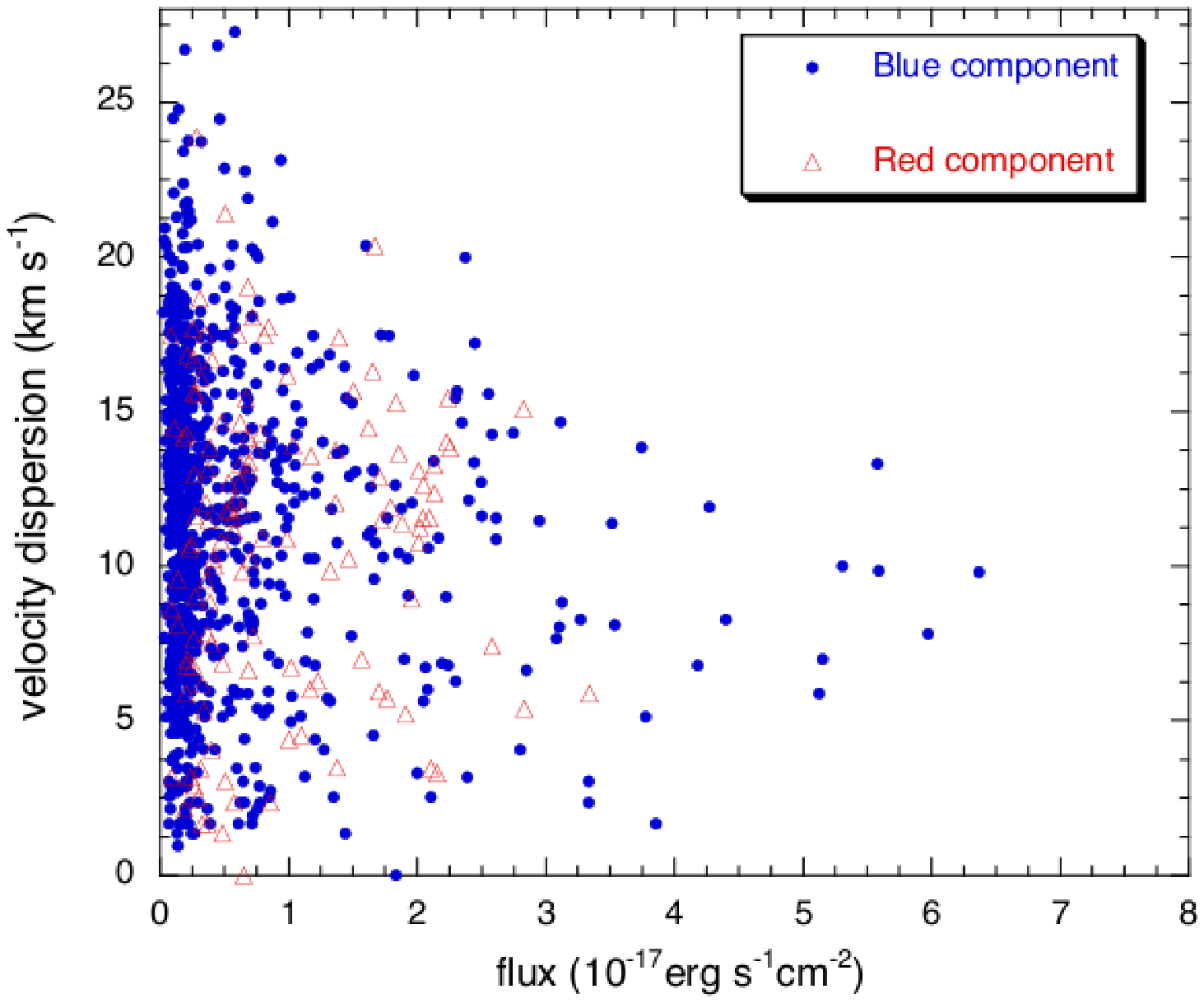}
\end{center}
\caption{\footnotesize  Top: a sketch of the kinematics of the gas suggested in the complex of NGC 6946. Velocity and velocity dispersion of the gas vs. relative distance along the complex. The relative flux of each component is also plotted. The model suggested for the kinematic of the gas, a large spreading bubble with small sub-bubbles on its wall at the same time, is indicated with the lines using the same color as the component of the gas that they represent.}
\label{fig:f5}
\end{figure}

\begin{figure*}
\begin{center}
\resizebox{\textwidth}{!}{
\includegraphics[width=0.45\textwidth,height=0.45\textwidth]{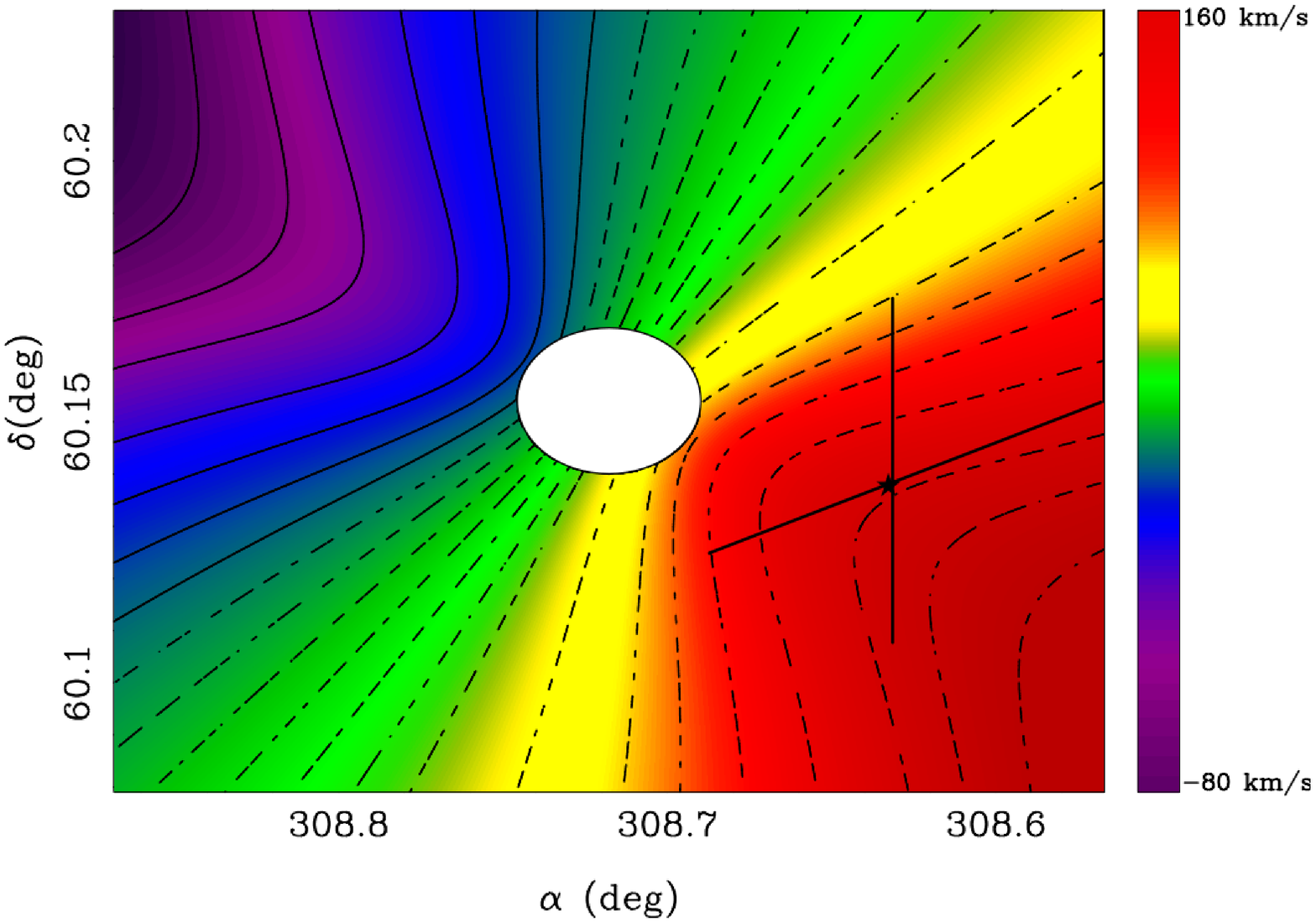}
\includegraphics[width=0.45\textwidth,height=0.5\textwidth]{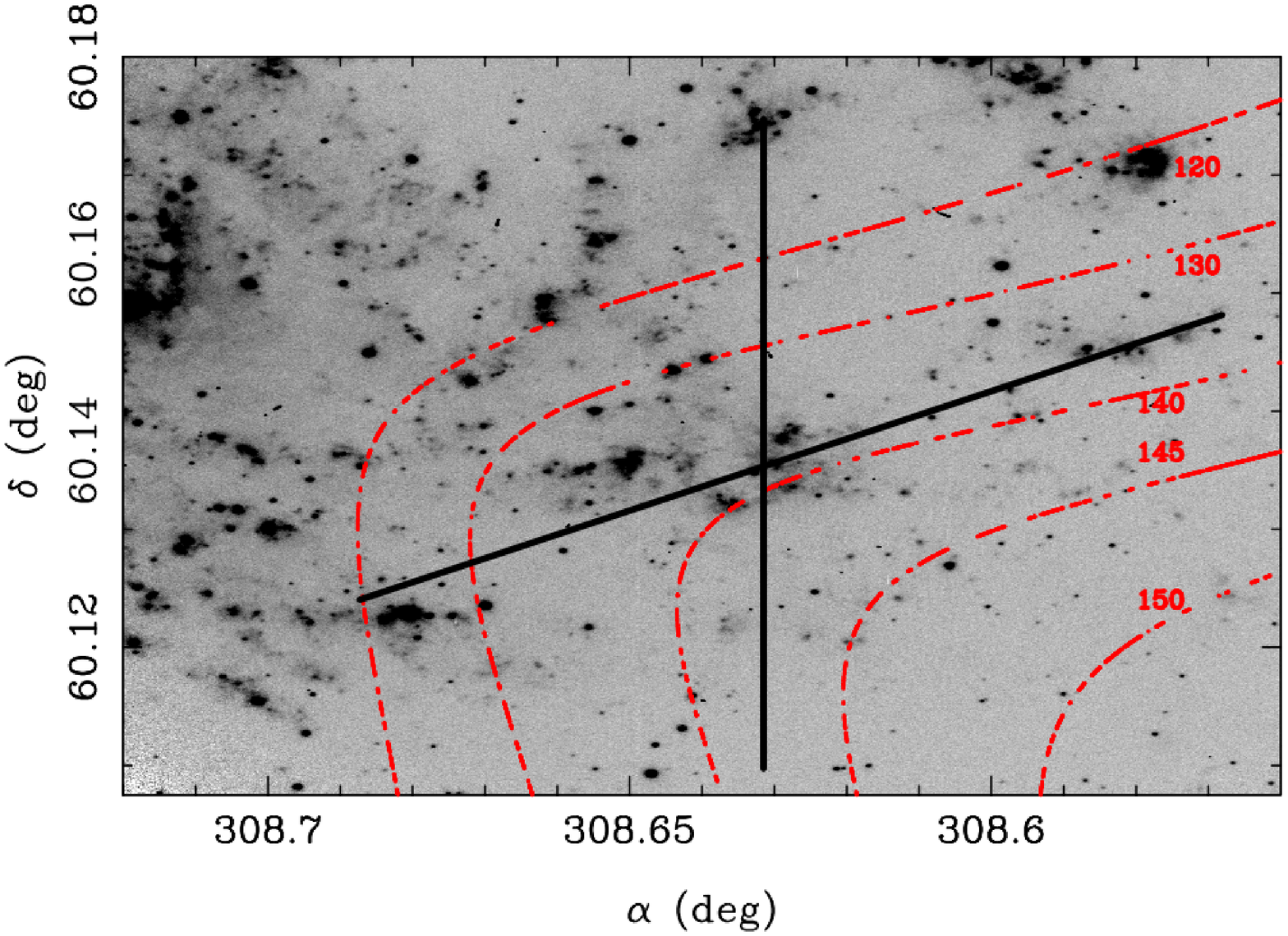}}
\caption{\footnotesize Rotation velocity map of NGC 6946.- (Left) two-dimensional model map obtained by fitting the rotation curve from  Blais-Ouellette et al. (2004) with the parametric model  $V_{pe}(r) = V_0 (1-e^{-r/r_{pe}}) (1+\alpha \, r/r_{pe})$ (Giovanelli \& Haynes 2002). Isophotes are represented with full lines for negative (approaching) velocities, and with dashed lines for positive (receding) velocities. The two slit positions are also represented and a star indicates the location of the SSC. The white circle in the middle of the map indicates the lack of  \Ha\ data at distances less than 50 arcsec from the galaxy center. (Right) \Ha\ image with five isovelocity contours overplotted as well as the two slit positions.}
\label{fig:f6}
\end{center}
\end{figure*}

These clumps of emission along the main shell of the complex are likely second generation star forming regions occurring in the shocked and highly compressed walls of the main shell, produced by the combined effects of previous generation stellar winds and supernova explosions. This shell is surrounded towards the inside and outside by very low  density gas. Thus when the second generation star forming regions start to expand due to their stellar winds it is probable that they easily break into the lower density surrounding medium and we would only be able to see either the approaching or receding sides (Elmegreen \& Lada 1977; Walborn \& Parker 1992, and references therein). 
There is no doubt that these regions are locally ionized by star clusters. This can be observed in the Gemini image, figure \ref{fig:f0} (or expanded in figure 2 of Efremov et al. 2007, or figure 1 of Larsen et al. 2002); the HII regions across the shell of the complex, corresponding to the small bubbles, clearly contain stellar clusters inside. 

An alternative explanation suggested by the large number of high velocity clouds (HVC) in NGC 6946 would be that these smaller bubbles are the result of impacts by some of these HVC against the main bubble of the complex. The possibility of a high-velocity cloud impact as a mechanism which might explain the formation of the complex is considered in Larsen et al. (2002).

Although the kinematic structure is quite symmetric overall, compatible with the idea of a bubble, it is not easy to explain the large size of the complex. Which mechanism caused this enormous complex is still under discussion. The idea of superexplosions has been put forward by Larsen et  al. (2002); a highly energetic explosion that swept the gas to the shell of a large bubble, subsequently triggering a second generation star formation in the shell. Formation of the younger stars might have been caused by such hypernova explosion ejected from the young super stellar cluster (YSSC; Efremov 2001), but even these explosions hardly explain the high energy required for such large bubble and the formation of the older stellar population in the complex or the YSSC. Comparison with HI intensity maps might help our understanding of the origin of this peculiar complex.

The bottom part of the top panel of figure \ref{fig:f5} shows the velocity dispersion of the blue component along the slit.
The velocity dispersion values have been corrected quadratically for instrumental broadening using the \Ha\  emission in the sky 
($\sigma_{sky} =24\pm5$ \kms). No clear tendency of the velocity dispersion along the slit is seen, although there is a hint of highs and lows anticorrelated with the line intensity. 

For an idealized shell in expansion, an anticorrelation is expected between the velocity dispersion and the \Ha\ intensity. This arises because in the center of the shell, as seen in projection, the integrated \Ha\ intensity is lower than in the border, while the velocity dispersion is highest in the center (given that the radial component of the expansion velocity is maximum there). Mu\~noz-Tu\~n\'on et al. (1996), Mart{\'{\i}}nez-Delgado et al. (2007) and Bordalo, Plana \& Telles (2009) have explored further this correlation in their studies of giant HII regions and of blue compact galaxies (cf. their figs. 2, 1 and 13, respectively). In the bottom panel of figure \ref{fig:f5} the expected trend of anticorrelation is clearly seen, confirming the overall geometry of an expanding shell.

The median value of $\sigma$ along the complex is  11 \kms, which corresponds with a hydrogen kinetic temperature of T$\sim$4800 K, through the expression
\begin{equation}
{\rm KE_{avg} = \overline{\Big[ \frac{1}{2}m v^2 \Big] } = \frac{3}{2}K_B T}
\end{equation}
where ${\rm K_B}$ is the Boltzmann constant, and we have used the velocity dispersion for ${\rm \overline{v^2}}$.

If we calculate the sound velocity in the cloud, assuming the ideal gas law, we have that
\begin{equation}
v_s = \sqrt{\frac{\gamma R T}{M}}
\end{equation}
where $\gamma$ is the adiabatic index, that for a monatomic ideal gas is 5/3; R is the molar gas constant; T the temperature, for which we use the value estimated above 4800 K;  and M the gas molar mass. Assuming a gas composed only of hydrogen, we obtain a value of $c_s = 8.2$ \kms. So the velocity dispersion measured is supersonic. If we assume a gas mass composition of ninety percent hydrogen and ten percent helium, the resulting sound velocity is even lower, $v_s = 7.1$ \kms, and the velocity dispersion is still supersonic.
Notice that although marginally supersonic, these values of the velocity dispersion are {\em spatially resolved} along each side of the shell, which implies that these are locally supersonic velocities in a scale of $\sim5.3$ pc (1 pixel). Supersonic line widths are one of the defining properties of giant extragalactic HII regions and starburst regions, and although their values are often measured of order or larger than 20 \kms, these correspond to much larger areas.  Bordalo, Plana \& Telles (2009) give a recent detailed analysis of supersonic motions in IIZw 40, but their resolved pixel is of order 100 pc; see their paper for a detailed discussion of supersonic motions from an observational point of view.

\subsection{Rotation curve of NGC 6946}\label{Sec3b}

To model the two-dimensional rotation of NGC 6946 we took the data from Blais-Ouellette et al. (2004), who measure the global \Ha\ rotation curves from Fabry-Perot data centered at \Ha. We apply the inverse procedure to compute  a two-dimensional rotation map, shown in figure \ref{fig:f6}. We compute the rotation map outside the central 50 arcsec radius, because the lack of \Ha\ emission in the central part of the galaxy makes the true kinematics in that central region difficult to retrieve. This does not affect the region of interest, the star forming complex that we study, which is located farther out, around 187.5$\arcsec$  (or 5 kpc), from the galaxy center (Larsen et al. 2002).
The reason for using the \Ha\ instead of the HI rotation curve (which would allow a complete map of the galaxy farther out), is due to the lower spatial resolution provided by the HI data which shows beam smearing (Blais-Ouellette et al. 1999; Swaters 1999; van den Bosch et al. 2000).

The rotation curve from figure 7 in Blais-Ouellette et al. (2004) is fitted using the parametric model of Giovanelli \& Haynes (2002),
\begin{equation}
V_{pe}(r) = V_0 (1-e^{-r/r_{pe}}) (1+\alpha \, r/r_{pe})
\label{eq:5}
\end{equation}
where $V_0$ regulates the overall amplitude of the rotation curve, $r_{pe}$ yields a scale length for the inner steep rise, and $\alpha$ sets the slope of the slowly changing outer part. After a least square fit of this model to the data, the parameters obtained are $[V_0, r_{pe}, \alpha]$ = [138$\pm$8 km s$^{-1}$, 40$\pm$5 arcsec, 0.037$\pm$0.008].
Let V(r) be the rotational velocity of an axially symmetric disk at a distance r from its center (r being the radial coordinate along the disk's projected major axis).
Let (x,y) be a set of Cartesian coordinates in the plane of the disk. If the disk is thin and it is inclined at an angle i (i = 90$^\circ$ for edge-on) to the line of sight, the observed component of the velocity, along the line of sight, that intercepts the disk at (x,y) is (Giovanelli \& Haynes 2002):
\begin{equation}
V_{||} = V(r) \frac{x}{\sqrt{x^2+y^2}}sin \, i + V_{nc}
\label{eq:6}
\end{equation}
where $V_{nc}$ accounts for non-circular motions and x is oriented along the disk's apparent major axis.
To obtain the observed component of rotational velocity we use the equation \ref{eq:6} above, ignoring the non-circular motions and adding the systemic velocity of NGC 6946, 145 km s$^{-1}$,
\begin{equation}
V_{||} = V(r) \frac{x}{\sqrt{x^2+y^2}}sin \, i + 145
\label{eq:7}
\end{equation}
The inclination angle used is $i \, = \, 38^\circ$ and the position angle of the line of nodes $60^\circ$ (Blais-Ouellette et al. 2004).

Once we have the rotation curve projected on our tangent disk, it is azimuthally expanded around the center of the galaxy to obtain the two-dimensional model map, shown in the left panel of figure \ref{fig:f6}. 
The location of the young SSC is marked by a star, and the two slit positions, at P.A. = 90$^\circ$ and 296$^\circ$, are represented by full lines. The full and dashed lines are the isophotes for negative (approaching) and positive (receding) velocities, respectively. The range of velocities shown spans from -80 km/s to 160 km/s. On the right hand side, the \Ha\ image of NGC 6946 around the complex is shown with the two slits and five model isovelocity contours.

\subsection{Size of the \Ha\ complex }\label{Sec3c}

\begin{figure}
\begin{center}
\includegraphics[width=0.6\textwidth,height=0.5\textwidth]{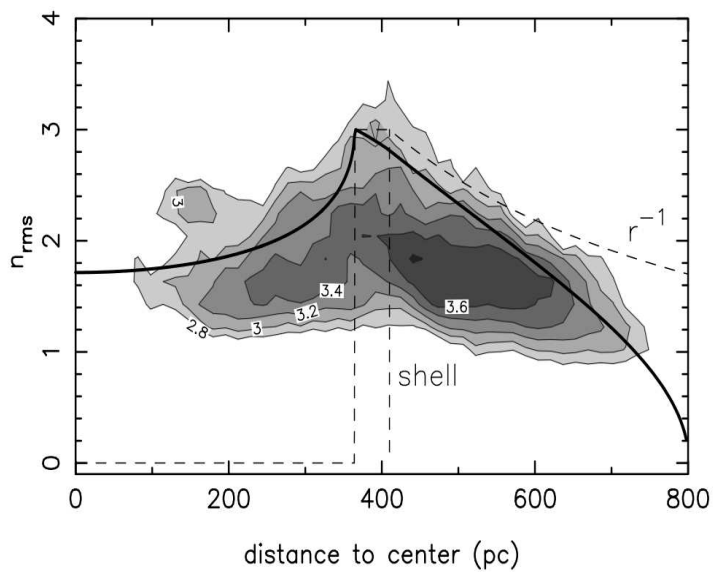} 
\caption{\footnotesize Root mean square (rms) density radial profile obtained from the \Ha\ image. The grayscale contours represent (in log units) the number of H$\alpha$ image pixels at that distance from the center of the bubble with a given value of the gas rms density (as obtained from the H$\alpha$ flux). The thin dashed line is the radial dependence of the gas density of an idealized shell, plus an $r^{-1}$ fall off law outside the shell. The thick line represents the integral of this radial density law, which gives a fair representation of the data, indicating that the ionized gas emission can indeed be approximated by a shell. 
If an extinction correction were applied to \Ha\ brightness of A$_B = 1$, the rms density values would increase by a 30\%.}
\label{fig:f7} 
\end{center}
\end{figure}

To obtain a value for the size of the shell radius, we use a method based on the characteristic radial profile of an emitting bubble. For a hollow shell of a given thickness seen in projection on the plane of the sky, the radial dependence of the observed flux from this shell has a characteristic form with a maximum at the radius of the shell, decreasing inwards to a minimum value at the shell center, and decreasing outwards to a vanishing flux value. Figure  \ref{fig:f7} shows the pixel-to-pixel radial dependence of the \Ha\ flux from the narrow band image.
Instead of the actual flux, we represent the equivalent root mean square (rms) density of the emitting gas\footnote{$n_{rms}(H_{\alpha}) \propto  \sqrt{n_p n_e} $, where $4 \pi j_{H_{\alpha}} / n_p n_e = 3.56 \times 10^{-25}$ (Case B at a temperature of $10^4$ K, Osterbrock \& Ferland 2006).  To calculate the emission coefficient $j_{H_{\alpha}}$, for an unresolved source and assuming that the source is uniformly emitting, the total luminosity is $L = 4 \pi j_{H_{\alpha}} V$.  Where V is the emitting volume, and the flux received at a distance {\it d} is $F_{H_{\alpha}} = L_{H_{\alpha}} /4 \pi d^2 = j_{H_{\alpha}} V / d^2$. }. Instead of plotting each pixel, for the sake of clarity we represent the number of pixels (in log units) with that value of  $n_{rms}(H_{\alpha})$ in a discrete grayscale. The thin dashed line represents the radial dependence of an idealized shell of inner radius 365 pc and 45 pc thickness, plus an $r^{-1}$ fall off density gradient outside the shell. The bold line is the integral of this radial density law projected on the plane of the sky. This resulting bold line provides a fair representation of the data. The \Ha\ emission in the complex is obviously much more structured as seen from the image, and in particular it has a significant contribution from the interstellar medium outside the idealized shell, which accounts for the emission outside the $365+45$ pc radius. In our simplified picture, this emission outside the idealized shell is represented by the r$^{-1}$ fall off density gradient. This is not meant to be a fit to the data, but rather an approximate idealized representation of the overall geometry of the region.

This method involves the uncertainty of having to choose the center of the complex. The central SSC (Larsen et al. 2002; Efremov et al. 2002) is taken initially as reference, being closer to the center of this spherical structure model. Different attempts were done, choosing the center of the bubble around in a small central region to the north-east of the SSC. The resulting radial distributions do not change significantly the pixel-density structure of the figure. 

The value of the inner radius that we obtain, 365 pc, is larger than the values given in the literature, of $\sim$300 pc (Larsen et al. 2002; Efremov et al. 2002), but it is somewhat smaller than the value expected observing the velocity-position plots, figure \ref{fig:f3}.

\subsection{Radial expansion velocities. Timescales.}\label{Sec3b1}

In order to calculate the actual velocities of the shell expansion we assume a bubble with spherical symmetry, and compute the radial velocities along the two slits on the projected geometry of the bubble, as explained below.

Neither of the slits runs through the center of the bubble, therefore the observed velocities are not exactly the radial velocities, since the slit does not determine a maximum circle. They are the projection of the corresponding component  of the radial velocity to line of sight, so the observed velocities are lower limits of the actual radial velocities. 

\begin{figure*}
\begin{center}
\resizebox{\textwidth}{!}{
\includegraphics[width=0.5\textwidth]{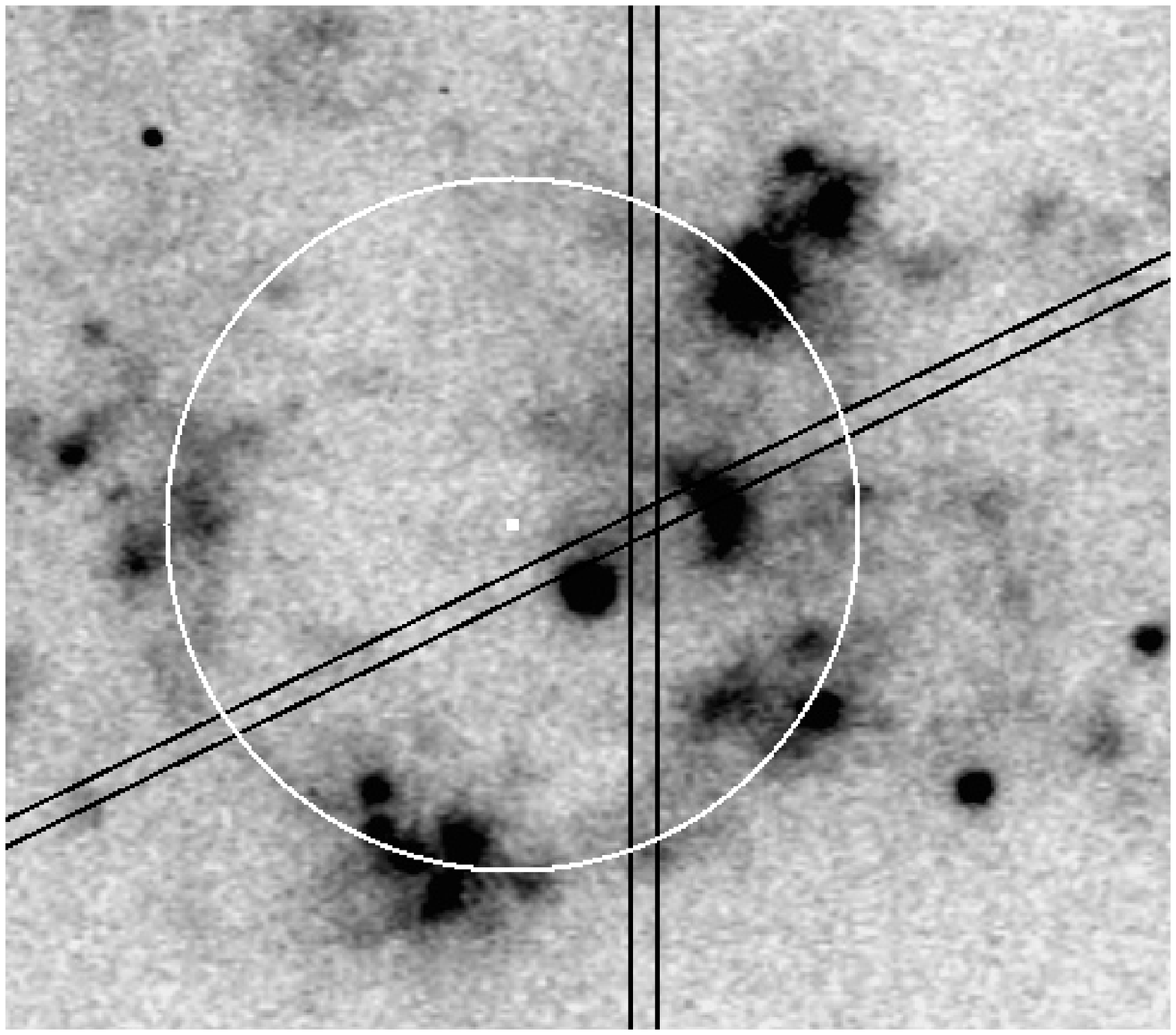} 
\includegraphics[width=0.5\textwidth]{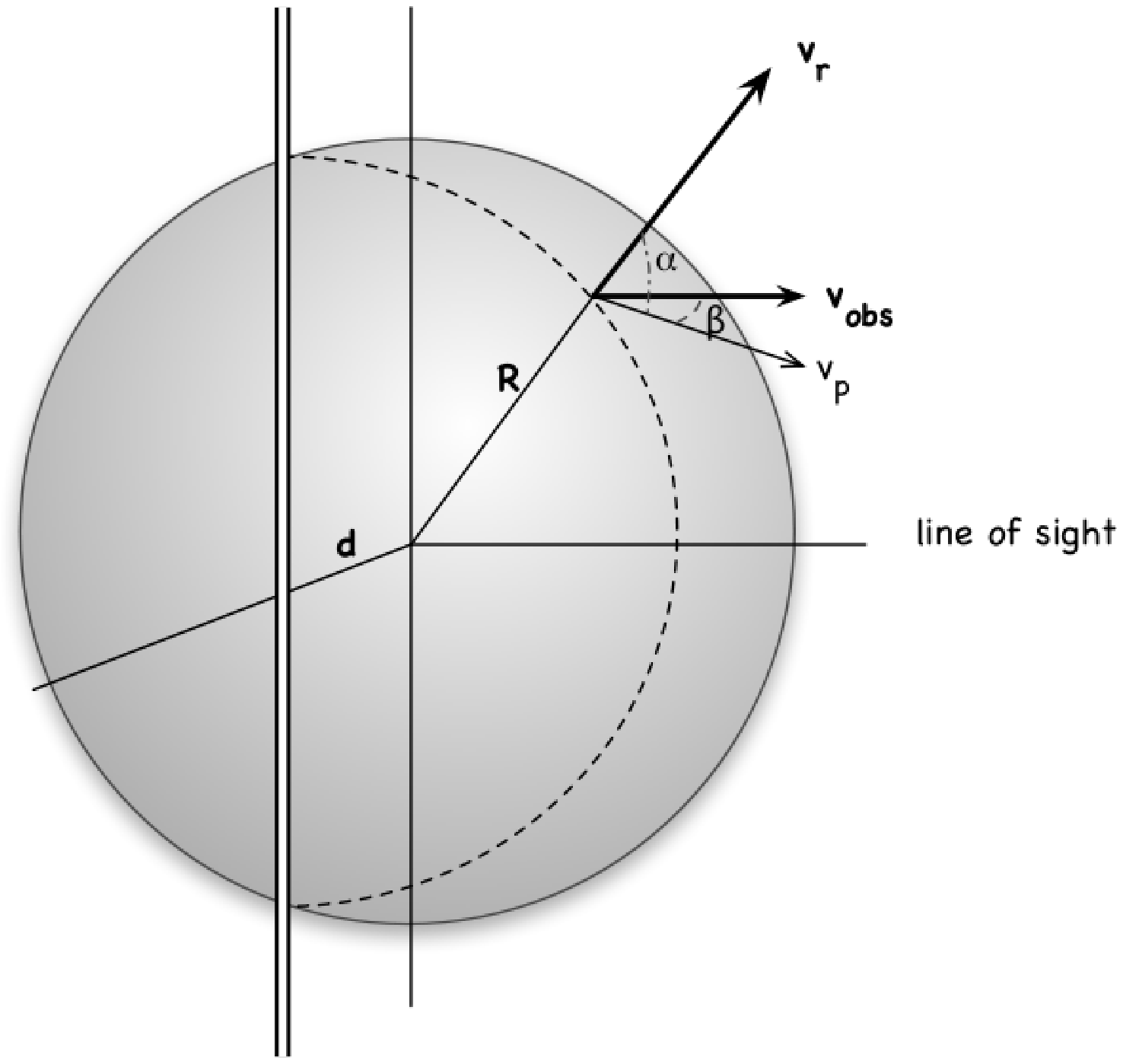} }
\caption{\footnotesize Left, an \Ha\ image of the complex with the slits and a representation of the inner radius resulting from the simplified shell model (section \ref{Sec3c}). Right, model of the bubble and slit geometry for the case of P.A. = 0$^\circ$.}
\label{fig:f8}
\end{center}
\end{figure*}

As the slit determines a circular section that does not contain the diameter of the sphere, the maximum velocity we can observe is the radial velocity multiplied by the cosine of the angle determined by the distance of the slit to the center of the sphere. This is, the observed velocity is then given by
\begin{equation}
v_{obs} = v_r \, cos\,  \alpha \, cos\,  \beta 
\label{vobs}
\end{equation}

\noindent where $v_r$ is the radial velocity of expanding gas, $\alpha$ the angle along the slit, and $\beta$ the angle from the spherical coordinates (figure \ref{fig:f8}), determined by the position of the slit. 

The maximum velocity along the slit is given by $v_{obs} =  v_r \, cos \beta $, corresponding to the projection of the radial velocity on the plane that contains the line of sight and is perpendicular to the slit, $\alpha \, = \, 0$. To obtain the expansion radial velocity from the observed value, we only have to determine the angle $\beta$ for $\alpha \, = \, 0$. This angle depends on both the distance of the slit to the center of the bubble, $d$, and also on the radius of the bubble, $R$, as follows
\begin{equation}
sin\, \beta = d/R
\label{beta}
\end{equation}

This simple relation is directly deduced from spherical coordinates, from the triangle formed in the plane containing the line of sight when $\alpha \, = \, 0$ (cf. figure \ref{fig:f8}). In section \ref{Sec3b} above we estimated both the bubble center and radius, R $\sim$ 365 pc, which inserted into these geometric relations, equations (\ref{vobs}) and (\ref{beta}), provide an estimation of the expansion velocity. 

For the slit at P.A. = 0$^\circ$, the distance of the slit to the bubble center is 118 pc, and the observed expansion velocities are around 35 $-$ 45 \kms , so from equations (\ref{vobs}) and (\ref{beta}), and taking $\alpha \, = \, 0$ to get the maximum value, the corresponding expansion velocities are $v_r \sim$ 37 $-$ 48 \kms. For P.A. = 296$^\circ$, the distance between the complex center and the slit is 66 pc, and the measured expansion velocities are $\sim$ 35 $-$ 50 \kms, giving true expansion velocities of $v_r \sim$ 36 $-$ 51 \kms. 

Using as the true expansion velocity, 37 \kms, it is possible to compute the expansion time of the main bubble through the simple formula $t_{dyn}$(Myr) $=$ R(pc)$/$v(\kms) $\approx$ 10 Myr.\footnote{$t_{dyn}$ is estimated assuming that the expansion velocity remains constant during the life of the bubble, giving just an order of magnitude for the age.}
This kinematic age is compatible with the age estimated for the stellar population of the SSC, $10-15$ Myr (Larsen et al. 2002), where the stellar winds and supernovae explosions would be the main source for the energetics of the shell expansion.

Similar calculations can be applied to a different part of the complex. In section \ref{Sec3a} above, we identified the small scale structures in the wall of the main expanding bubble as the bubbles corresponding to the second generation of star formation, produced by the compression of the interstellar medium by the expansion of the main shell (cf. Larsen 2002). If we take one of these small bubbles on the wall of the main shell, measure its radius, r $\sim$ 80 pc, and its observed expansion velocity (with respect to the expansion velocity of the main shell) of $\sim20$ \kms, giving then a $t_{dyn} \sim$ 4 Myr. 
The expansion time of this small bubble is compatible with the stellar ages of this second generation of star formation, given that  Efremov et al. (2002) find an age of 4 Myr for the stellar population in the arc shaped structure to the northwest of the central SSC.

\section{Diagnostic Diagrams}\label{Sec4}

Plots of emission line ratios are developed in this section with the aim to determine the main mechanism of excitation of the ionized gas. The structure and stellar evolution of this star forming complex encompasses a suit of ionization and excitation processes, including photoionization by hot stars and shocks by stellar winds and supernova explosions amongst other, and here we explore whether with these data sets it is possible to distinguish them. 

Baldwin, Philips \& Terlevich (1981; BPT) researched how to classify empirically extragalactic objects based on their predominant excitation mechanism, from several combinations of pairs of easily measurable emission-line ratios. They separate these mechanisms into four categories: gas ionized by hot stars of type O and B, as in HII regions or star forming regions; planetary nebulae, photo-ionized by stars hotter than type O; photo-ionization by a non-thermal continuum following a power law, as in AGNs; and gas heated by shock waves. 

BPT found that these four groups can be efficiently separated by plotting [NII]$\lambda$6584/\Ha\ vs. [OIII]$\lambda$5007/H$\beta$, and [OII]$\lambda$3727/[OIII]$\lambda$5007 vs. [OIII]$\lambda$5007/H$\beta$, [NII]$\lambda$6584/\Ha, or [OI]$\lambda$6300/\Ha. 

These diagrams have been used extensively and their utility for classification has been widely demonstrated. They were used also as abundance diagnostics in emission nebulae (Sabbadin, Minello \& Bianchini 1977; D'Odorico 1978; Dopita 1977, 1978). 

Looking for the best choice among all the possible emission line ratios, Veilleux \& Osterbrock (1987) gave some criteria that include choosing strong nearby lines which are not blended; preferably a ratio of a forbidden to an HI Balmer. The line ratios satisfying these criteria are: [OIII]$\lambda$5007/H$\beta$, [NII]$\lambda$6583/\Ha, and [SII]($\lambda$6716+$\lambda$6731)/\Ha. 

\subsection{Computing the emission line ratios}\label{Sec3c1} 

\begin{figure}
\begin{center}
\includegraphics[width= 0.6\textwidth]{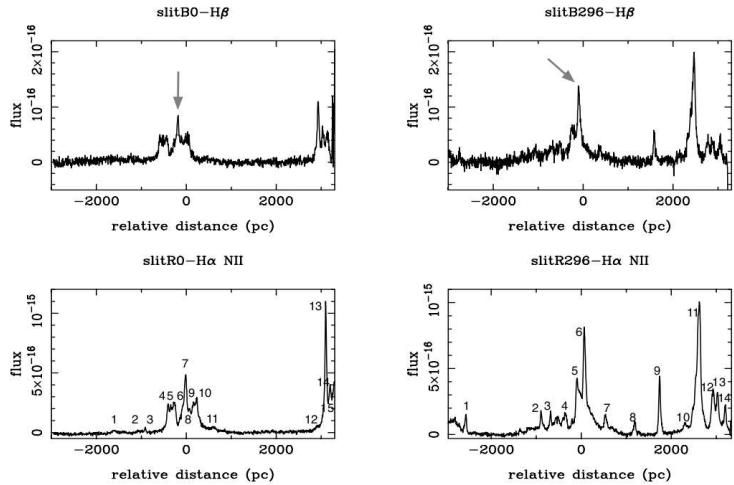} 
\caption{\footnotesize Spatial cuts along the slit at P.A. = 0$^\circ$ and P.A. = 296$^\circ$ for the blue and red spectral ranges. Abscissa is in parsec, and ordinate represents flux in erg cm$^{-2}$ s$^{-1}$\AA$^{-1}$. The location of the complex is marked with arrows in the cuts around H$\beta$; numbers indicate the locations were spectra were summed and extracted for the diagnostic diagrams. }
\label{fig:f9}
\end{center}
\end{figure}

In order to achieve enough signal-to-noise in [OIII] and H$\beta$, which are faint in our spectra, we have extracted the spectra adding a few pixels around the brightest regions.  Figure \ref{fig:f9} shows spatial cuts of spectra along P.A. = 0$^\circ$ and P.A. = 296$^\circ$, each one centered at the wavelengths around \Ha\ and H$\beta$. The whole length of the slit is shown, and the location of the star forming complex is marked by an arrow in the cuts around H$\beta$. In the cuts centered at \Ha, numbers indicate the locations from where the spectra have been extracted. We used DIPSO to fit  \Ha, [NII], and [SII] from the red spectra, and H$\beta$, and [OIII]  from the blue spectra, at both slit P.A., and then computed the flux ratios $log (I_{[OIII] \lambda 5007} / I_{H\beta})$ , $log (I_{[NII] \lambda 6584} / I_{H\alpha})$  and $ log(I_{[SII] (\lambda6716+\lambda6731)} / I_{H\alpha})$.

Unfortunately the emission in the blue spectra is still very faint even after the coadded extraction, and there are very few points remaining with enough signal to fit the H$\beta$ and [OIII] lines. In this circumstance we decided to use the diagnostic diagram log([SII]6717+6731/\Ha) versus log([NII]6584/\Ha). We use two types of measurements to construct the diagnostic diagram, pixel by pixel fluxes along the slit and coadded extractions around the brightest knots. The models used to analyse these ratios and to determine the ionization mechanism involved are described below.

\begin{figure*}
\resizebox{\textwidth}{!}{
\includegraphics[width= 0.5\textwidth,height=8cm]{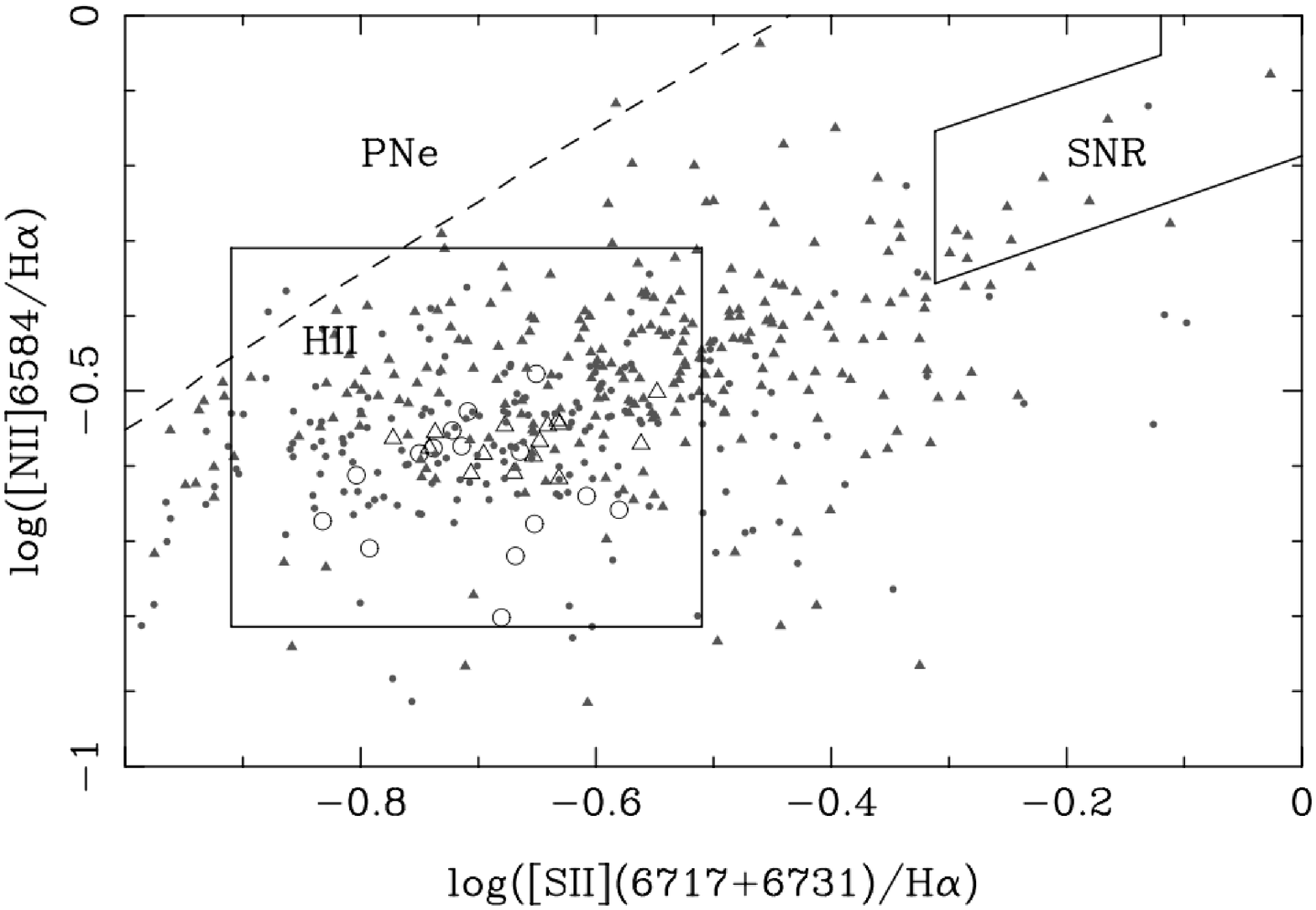} 
\includegraphics[width= 0.5\textwidth,height=8cm]{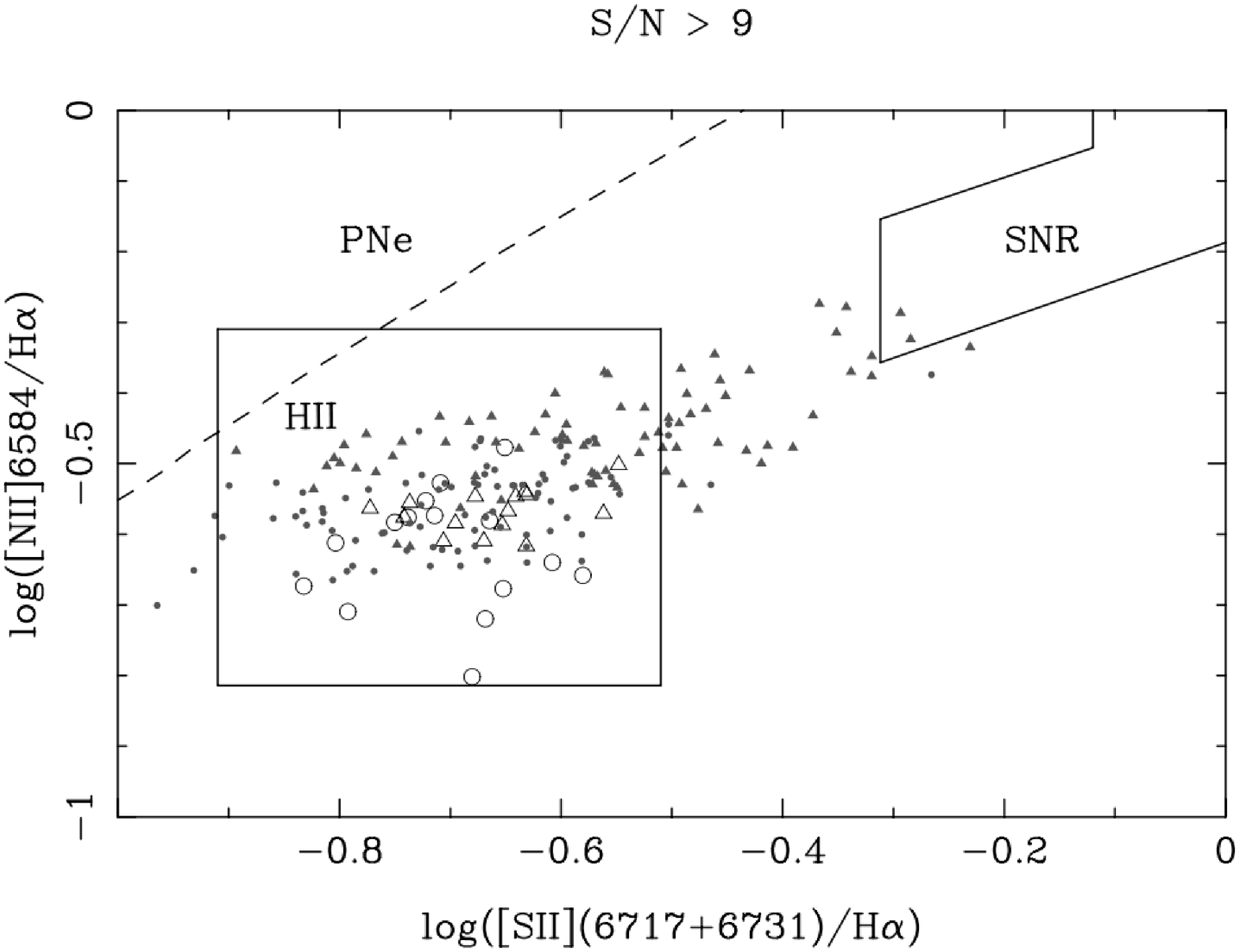} 
}
\caption{\footnotesize Diagnostic diagram of the brightest emission line ratios in the red spectral range. The left panel shows all the ratios measured along the slit, while the right panel shows only those with signal/noise $ > 9\sigma_f$. The small filled symbols represent the pixel by pixel measurements along the slit across the star forming complex, while the large open symbols represent the coadded extractions around the brightest knots (circles for P.A.=0º and triangles for P.A.=296º).}
\label{fig:f10}
\end{figure*}

I.- In  figure \ref{fig:f10} we use the models by Sabbadin, Minello \& Bianchini (1977; SMB77) as regions separating main groups of ionized regions dominated by different ionization and excitation mechanisms. SBM77 introduced the electronic-excitation density diagrams with the aim to understand the nature of the nebulae S176. These diagrams compare the emission line ratios \Ha/[NII], [SII]$\lambda\lambda$6717/6731 and \Ha/[SII] observed in supernova remnants (SNR), planetary nebulae (PNe) and HII regions. These nebulae are supposed to be in different regions in the diagrams due to the different ionization and excitation processes governing their nebular structure. The model is adapted to the ratios that we measure in this work. SBM77 model is still widely used, e.g., L\'opez-Mart\'in et al. (2002); Riesgo \& L\'opez (2006); Phillips \& Cuesta (1999).

SMB77 model locates the complex  in the region associated to photoionization by hot massive stars. Most of the points are inside or close to the HII region box, as it can be seen in the top panel of figure \ref{fig:f10}. A fraction of the points lie outside the HII region box; those few points indicating low values of the two ratios may correspond to higher excitation regions. More significantly, the points corresponding to higher values of the two ratios, located in the transition zone between HII and SNR,  could indicate the presence of shocked gas, which would be represented by larger [SII]/\Ha\ and [NII]/\Ha\ ratios. 
This result is suggestive but tentative, because only a few data points remain with fluxes above 9$\sigma_f$ outside the HII box. The overall ionization structure of a star forming complex is dominated by the hot massive stars, while the impact in the ionization structure by the SNR and PNe in these giant complexes is less readily ascertained in the diagnostics diagrams (Martin 1997). Thus, this result is promising but higher spatial resolution and signal-to-noise spectra are needed to confirm it. Notice that this is consistent with the supersonic velocity dispersion values measured across the shell.

\begin{figure}
\begin{center}
\includegraphics[width= 0.6\textwidth]{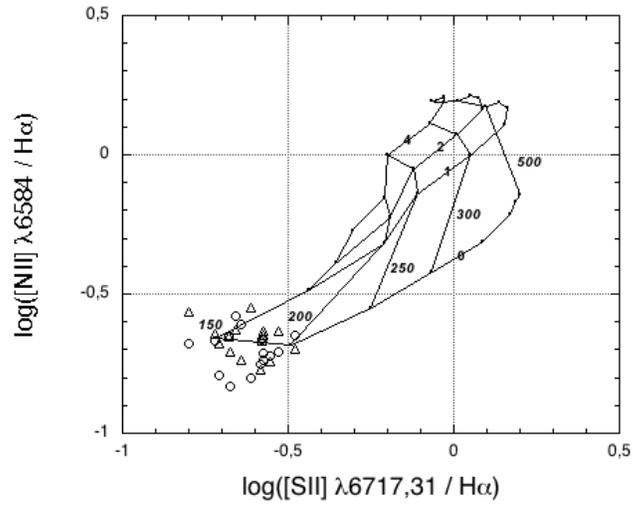}
\caption{\footnotesize Comparison of the emission line ratios log([NII]$\lambda$6584/\Ha) vs. log([SII]$\lambda$6717,6731/\Ha) with the DS96 high velocity shocks grid of models, with shock velocities in the range 150 $\leq$ V$_S$ $\leq$ 500 \kms , and magnetic parameter 0 $\leq$ B/n$^{1/2}$ $\leq$ 4 $\mu$G cm$^{-3/2}$. The circles and triangles represent ratios from the spectrum at P.A. = 0$^\circ$ and at P.A. = 296$^\circ$ respectively. }
\label{fig:f11}
\end{center}
\end{figure}

II.- In figure \ref{fig:f11} we use the model by Dopita \& Sutherland (1996; DS96). This is a high velocity radiative shock model, including the magnetic pressure as support of the photoionization/recombination zone. It is a more complete and elaborate model, and it can explain some objects which cannot be understood in terms of the simpler  photoionization models with a central source. The basic physical idea is that a high velocity radiative shock in the interstellar medium is an important source of ionizing photons. These photons are produced in the cooling of hot plasma after the shock, and they may spread encountering the pre-shock gas and forming a precursor HII region. We use this model to look for some clues of shocks within the complex. Data to represent the grid are obtained from table 8 in DS96, paper I. This is a low-density steady-flow model,  the shock velocities lie in the range 150 $\leq$ V$_S$ $\leq$ 500 \kms , and the magnetic parameter 0 $\leq$ B/n$^{1/2}$ $\leq$ 4 $\mu$G cm$^{-3/2}$.

When comparing the emission line ratios with the grid of DS96 model, figure \ref{fig:f11}, data points are located in zone of shocks with low velocities and magnetic parameter. Better S/N data would be required to make a more detailed study of the presence of shocks and their role. Efremov et al. (2002) conclude that the intensities of Balmer, [NII] and [SII] emission lines within and around the complex indicate that shock excitation makes a contribution to the emission from the most energetic region.


\section{Summary} \label{Sec5}

The gas-star super-complex located in the galaxy NGC 6946 is an excellent laboratory for the study of the interaction between the massive star formation and the gas cloud where the stars originated. The presence of a SSC with a mass around 10$^6$ M$\odot$ and an age younger than 20 Myr appears to be the main source of ionizing photons able to generate an \Ha\ bubble near 730 pc in diameter. 
Previous studies of this star forming complex indicate its shell geometry (Larsen \& Ritchler 1999; Efremov 1999; Elmegreen et al. 2000; Efremov et al. 2002, 2007; Larsen et al. 2002), and here we carry out a comprehensive analysis of the kinematical and photometric parameters of a shell in expansion.
Moreover, in this study we obtain for the first time the large scale velocity field of both bubble sides, the approaching and receding walls, which shows a complicated pattern with sub-bubbles forming at the walls of the largest structure, in a similar scheme to the formation 'a la Huygens', where some points on the surface of the previous bubble become the formation center of the second generation bubbles.
The estimated ages (Elmegreen et al. 2000; Larsen et al. 2002) of the super cluster and of the younger stellar population forming an arch like structure at the Northwest side of the bubble centre are compatible with this scenario. 
This spectroscopic  analysis provides detailed information about the spatial distribution along the bubble of several variables concerning the physical state of the gas such as gas density, velocity dispersion, and line ratios, making this object the best available probe for testing dynamical and physical models of super bubble generation. 

The radius calculated for the main bubble is 365 pc, somewhat larger than the values given in the literature at the moment. The radial expansion velocity of the complex is calculated in the range 37 \kms\ to 50 \kms. Furthermore, with the lower limit of the radial velocities, $v_r=37 $ \kms, an estimation of the (dynamical) expansion time for the complex  is about 10 Myr. This age is consistent with the stellar ages within the central part of the complex. This fact supports  the hypothesis for the origin of the large bubble due to a sequence of supernova explosions, which would provide enough energy for such a large bubble. 
 
The diagnostic diagram from the emission line ratios show that most of the ionization arises from energetic photons from massive stars, while there are indications that some locations may be ionized by low velocity shocks from stellar winds and/or supernova explosions, consistent with the supersonic velocity dispersion measured throughout the shell.

\section*{acknowledgment}

We are very grateful to Yuri N. Efremov, Bruce G. Elmegreen and Nate Bastian for advice and discussion during the realization of this work. Thanks to Guillermo Tenorio Tagle for a discussion of the expansion of the nebula during the photoionization phase. And finally, we are deeply grateful to the referee for constructive comments and valuable advice.

This work is part of the PhD dissertation of M. Carmen Sánchez-Gil, funded by the Spanish Ministerio de Educación y Ciencia, under the FPU grant AP-2004-2196. We acknowledge financial support
from Spanish MICINN through grants AYA2007-64052 and AYA2007-64712 and from Consejer\'{\i}a de Educaci\'on y Ciencia (Junta de Andaluc\'{\i}a) through TIC-101, TIC-4075 and TIC-114.

%


\begin{table*}
\caption{Observational configuration of different spectroscopic studies}
\begin{center}
\resizebox{\textwidth}{!}{
\begin{tabular}{c c c c c c}
\hline
Instrument \& Detector & Wavelength range (\AA) \quad&Exposure time (s) \quad& Dispersion (\AA / pixel) \quad& Pixel scale (\arcsec / pixel) &Seeing (\arcsec) \\ 
 \hline
& && & &\\
 Multi-Pupil Fiber Spectrogragh, 6m BTA. {\bf [1]}&6140 $-$ 7120 & 3$\times$900 &0.76 & 0.14 &1.2, 1.5 \\
  CCD EEV42-40 (2048$\times$2048 pix)& && &&\\

6m SAO RAS telescope {\bf [2]}& 6015 $-$ 7250& 1,2$\times$1800 & 1.2 & 0.39&1.4,2.7\\
Photometrics CCD-detector (1024$\times$1024 pix) & && & &\\

Keck-I 10m telescope {\bf [3]}& 6220 $-$ 8550& 4$\times$3300 & 0.048 & --&0.9\\
D1 decker (14\arcsec $\times$ 1\arcsec.15 slit) & && & &\\

ISIS at William Herschel Telescope & 6034 $-$ 7088 & 3$\times$1800 &0.23 & 0.2 &1.0\\
MARCONI2 CCD (4096$\times$2048 pix)& && & &\\
\hline
\end{tabular}
}
\vspace{0.5 cm}
{\footnotesize 
\begin{description}
\item [NOTES.-] Values given within the table are only related to \Ha\ range data, for the analysis of the velocity field. 
\item [[1]] Efremov et al. 2007. 
\item [[2]] Efremov et al. 2002.
\item [[3]] Efremov et al. 2002. But limited to a full extent of 14\arcsec along the slit, with a position angle of -10$^\circ$, close to the SSC location.

\end{description}}
\end{center}
\label{table1}
\end{table*}%

\begin{table*}
\caption{Journal of observations}
\begin{center}
\resizebox{0.8\textwidth}{!}{
\begin{tabular}{c c c c c c }
\hline
Date & P.A. (deg) & Wavelength range (\AA) &Exposure time (s) \quad & Slit (\arcsec)& Airmass\\ 
 \hline
21.08.2003& 0& 6034$-$7088&2$\times$1800 $+$ 1200 & 1.0&1.23\\
21.08.2003&296& 6034$-$7088& 3$\times$1800& 1.0&1.17\\
\hline
\end{tabular}}
\end{center}
\label{table2}
\end{table*}%

\begin{table*}
\caption{Summary of the emission line fitting}
\begin{center}
\resizebox{0.8\textwidth}{!}{
\begin{tabular}{l  c c  c c c c c c}
\hline
slit & \multicolumn{2}{c}{Velocity$^a$}& \multicolumn{2}{c}{FWHM$^a$} & \multicolumn{2}{c}{EW$^b$} &\multicolumn{2}{c}{Flux$^c$} \\
& \makebox[0.8 cm]{$\overline{V_{obs}}$} & \makebox[0.8 cm]{$\sigma_{V}$} & \makebox[0.8 cm]{$\overline{W_{obs}}$} &  \makebox[0.8 cm]{$\sigma_{W}$} &
\makebox[0.8 cm]{$\overline{EW}$} & \makebox[0.8 cm]{$\sigma_{EW}$} &\makebox[0.8 cm]{$\overline{F}$}& \makebox[0.8 cm]{$\sigma_{F}$} \\
\hline
&&&&&&&&\\
\makebox[1.3cm][r]{$^{(i)}$}&106& 25& 27 & 12&9 & 9&0.99 &1.08 \\
P.A. = 0$^{\circ}$ &&&&&&&&\\
\makebox[1.3cm][r]{$^{(ii)}$}&148 & 29&25 & 11&12 & 11& 1.19 & 0.89  \\
\hline
&&&&&&&&\\
\makebox[1.4cm][r]{$^{(i)}$}&120 &24&24 & 10&10 & 15&0.82  & 1.25\\
P.A. = 296$^{\circ}$&&&&&&&&\\
\makebox[1.4cm][r]{$^{(ii)}$}&169 & 20&23 & 9&7 & 7&0.6 &0.53 \\
\hline
\end{tabular}
}
\end{center}

$^a$ \kms; $^b$ \AA; $^c$ 10$^{-17}$erg~cm$^{-2}$~s$^{-1}$~\AA $^{-1}$ \\

NOTE.- Mean ($\overline{X}$) and standard deviation ($\sigma_X$) of the variables obtained from the fits for the blue ({\it i}) and red ({\it ii}) components of the gas, for both slits. Values correspond only to the central part of the spectra, where the complex is located.
\label{table3}
\end{table*}%


\begin{thebibliography}{40}
 
\bibitem[Alfaro \& El{\'{\i}}as(2006)]{2006LNEA....2..111A} Alfaro, E., \& El{\'{\i}}as, F.\ 2006, Lecture Notes and Essays in Astrophysics, vol.~2, p.~111-126., 2, 111 

\bibitem[Alfaro et al.(2007)]{2007acag.conf..437A} Alfaro, E.~J., Garc\'\i a-Lorenzo, B., Efremov, Y.~N., \& Mu\~noz-Tu\~n\'on, C.\ 2007, Astrophysics and Cosmology After Gamow, 437 

\bibitem{1981PASP...93....5B} Baldwin, J.~A., Phillips, M.~M., \& Terlevich, R.\ 1981, \pasp, 93, 5 

\bibitem[Beck(1991)]{1991A&A...251...15B} Beck, R.\ 1991, \aap, 251, 15 

\bibitem[Bekki \& Chiba(2006)]{2006ApJ...637L..97B} Bekki, K., \& Chiba, M.\ 2006, \apjl, 637, L97 

\bibitem[Blais-Ouellette et al.(1999)]{1999AJ....118.2123B} Blais-Ouellette, S., Carignan, C., Amram, P., \& C{\^o}t{\'e}, S.\ 1999, \aj, 118, 2123 

\bibitem[Blais-Ouellette et al.(2004)]{2004A&A...420..147B} Blais-Ouellette, S., Amram, P., Carignan, C., \& Swaters, R.\ 2004, \aap, 420, 147 

\bibitem[Boomsma(2007)]{2007PhDT.........1B} Boomsma, R.\ 2007, Ph.D.~Thesis,  

\bibitem[Boomsma et al.(2008)]{2008A&A...490..555B} Boomsma, R., Oosterloo, T.~A., Fraternali, F., van der Hulst, J.~M., \& Sancisi, R.\ 2008, \aap, 490, 555 

\bibitem[Bonnarel et al.(1986)]{1986A&AS...66..149B} Bonnarel, F., Boulesteix, J., \& Marcelin, M.\ 1986, \aaps, 66, 149 

\bibitem[Bonnarel et al.(1988)]{1988A&A...189...59B} Bonnarel, F., Boulesteix, J., Georgelin, Y.~P., Lecoarer, E., Marcelin, M., Bacon, R., \& Monnet, G.\ 1988, \aap, 189, 59 

\bibitem[Bordalo et al.(2009)]{2009ApJ...696.1668B} Bordalo, V., Plana, H., \& Telles, E.\ 2009, \apj, 696, 1668

\bibitem[Dodorico(1978)]{1978MmSAI..49..485D} Dodorico, S.\ 1978, Memorie della Societa Astronomica Italiana, 49, 485 

\bibitem[Dopita(1977)]{1977ApJS...33..437D} Dopita, M.~A.\ 1977, \apjs, 33, 437 

\bibitem[Dopita(1978)]{1978ApJS...37..117D} Dopita, M.~A.\ 1978, \apjs, 37, 117 

\bibitem{1996ApJS..102..161D} Dopita, M.~A., \& Sutherland, R.~S.\ 1996, \apjs, 102, 161 

\bibitem[Efremov(1999)]{1999AstL...25...100} Efremov, Y.~N.\ 1999, 
Astronomy Letters, 25, 100

\bibitem[Efremov(2001)]{2001ARep...45..769E} Efremov, Y.~N.\ 2001, Astronomy Reports, 45, 769 

\bibitem[Efremov et 
al.(2002)]{2002A&A...389..855E} Efremov, Y.~N., et al.\ 2002, \aap, 389, 855 

\bibitem{2004A&AT...23...95E} Efremov, Y.N., Alfaro, E., Hodge, P., Larsen, S., \& Mu\~noz-Tu\~n\'on, C.\ 2004, Astronomical and Astrophysical Transactions, 23, 95 

\bibitem[Efremov et al.(2007)]{2007MNRAS.382..481E} Efremov, Y.~N., et al.\ 2007, \mnras, 382, 481 

\bibitem[Elmegreen \& Lada(1977)]{1977ApJ...214..725E} Elmegreen, B.~G., \& Lada, C.~J.\ 1977, \apj, 214, 725 

\bibitem[Elmegreen et al.(2000)]{2000ApJ...535..748E} Elmegreen, B.~G., Efremov, Y.~N., \& Larsen, S.\ 2000, \apj, 535, 748 

\bibitem[Giovanelli \& Haynes(2002)]{2002ApJ...571L.107G} Giovanelli, R., \& Haynes, M.~P.\ 2002, \apjl, 571, L107 

\bibitem{2006 ApJ...647...1018}Heald, G. H., Rand, R. J., Benjamin, R. A., Bershady, M. A., 2006, \apj, 647, 1018

\bibitem[Hodge(1967)]{1967PASP...79...29H} Hodge, P.~W.\ 1967, \pasp, 79, 29 

\bibitem[Kamphuis(1993)]{1993sfgi.conf..105K} Kamphuis, J.\ 1993, Star Formation, Galaxies and the Interstellar Medium, 105 

\bibitem{1999A&A...345...59L} Larsen, S.~S., \& Richtler, T.\ 1999, \aap, 345, 59 

\bibitem[Larsen et al.(2001)]{2001ApJ...556..801L} Larsen, S.~S., Brodie, J.~P., Elmegreen, B.~G., Efremov, Y.~N., Hodge, P.~W., \& Richtler, T.\ 2001, \apj, 556, 801 

\bibitem{2002ApJ...567..896L} Larsen, S.~S., Efremov, Y.~N., Elmegreen, B.~G., Alfaro, E.~J., Battinelli, P., Hodge, P.~W., \& Richtler, T.\ 2002, \apj, 567, 896 

\bibitem{2006MNRAS.368L..10L} Larsen, S.~S., Origlia, L., Brodie, J.~P., \& Gallagher, J.~S.\ 2006, \mnras, 368, L10

\bibitem[L{\'o}pez-Mart{\'{\i}}n et al.(2002)]{2002A&A...388..652L} L{\'o}pez-Mart{\'{\i}}n, L., et al.\ 2002, \aap, 388, 652 

\bibitem[Martin(1997)]{1997ApJ...491..561M} Martin, C.~L.\ 1997, \apj, 491, 561 

\bibitem[Mart{\'{\i}}nez-Delgado et al. (2007)]{MD2007} Mart{\'{\i}}nez-Delgado, I., Tenorio-Tagle, G., Mu{\~n}oz-Tu{\~n}{\'o}n, C., Moiseev, A.~V., \& Cair{\'o}s, L.~M.\ 2007, \aj, 133, 2892 


\bibitem[Mu\~noz-Tu\~n\'on et al. (1996)]{MT1996} Mu\~noz-Tu\~n\'on, C., Tenorio-Tagle, G., Casta\~neda, H.~O., \& Terlevich, R.\ 1996, \aj, 112, 1636 

\bibitem[Oke(1990)]{1990AJ.....99.1621O} Oke, J.~B.\ 1990, \aj, 99, 1621 

\bibitem[Osterbrock, D. E. \& G. Ferland (2006)]{} Osterbrock, D. E. \& G. Ferland. 2006. Astrophysics of Gaseous Nebulae and Active Galactic Nuclei, Science University Press.

\bibitem[Phillips \& Cuesta(1999)]{1999AJ....118.2919P} Phillips, J.~P., \& Cuesta, L.\ 1999, \aj, 118, 2919 

\bibitem{2006RMxAA..42...47R} Riesgo, H., \& L{\'o}pez, J.~A.\ 2006, Revista Mexicana de Astronomia y Astrofisica, 42, 47

 \bibitem[Rozas et al.(2000)]{2000A&A...354..823R} Rozas, M., Zurita, A., \& Beckman, J.~E.\ 2000, \aap, 354, 823 

\bibitem[Rozas et al.(1999)]{1999A&AS..135..145R} Rozas, M., Zurita, A., Heller, C.~H., \& Beckman, J.~E.\ 1999, \aaps, 135, 145 

\bibitem{1977A&A....60..147S} Sabbadin, F., Minello, S., \& Bianchini, A.\ 1977, \aap, 60, 147 

\bibitem[Sofue(1997)]{1997PASJ...49...17S} Sofue, Y.\ 1997, \pasj, 49, 17 

\bibitem[Swaters(1999)]{1999ASPC..182..369S} Swaters, R.\ 1999, Galaxy Dynamics - A Rutgers Symposium, 182, 369 

\bibitem[Tenorio-Tagle et al.(2005)]{2005ApJ...620..217T} Tenorio-Tagle, G., Silich, S., Rodr{\'{\i}}guez-Gonz{\'a}lez, A., \& Mu{\~n}oz-Tu{\~n}{\'o}n, C.\ 2005, \apj, 620, 217

\bibitem[van den Bosch et al.(2000)]{2000AJ....119.1579V} van den Bosch, F.~C., Robertson, B.~E., Dalcanton, J.~J., \& de Blok, W.~J.~G.\ 2000, \aj, 119, 1579 

\bibitem[Veilleux \& Osterbrock(1987)]{1987ApJS...63..295V} Veilleux, S., \& Osterbrock, D.~E.\ 1987, \apjs, 63, 295 

\bibitem[Walborn 
\& Parker(1992)]{1992ApJ...399L..87W} Walborn, N.~R., \& Parker, J.~W.\ 1992, \apjl, 399, L87 


\end{thebibliography}
\end{document}